\documentclass[aps,prd,nofootinbib,onecolumn,superscriptaddress,preprintnumbers,balancelastpage,longbibliography, 11pt]{revtex4-1}
\usepackage{amsmath,amssymb,mathtools,bm}
\usepackage{graphicx, color, hepunits}
\usepackage[dvipsnames]{xcolor}
\usepackage{float}
\usepackage{filecontents}
\usepackage{multirow}
\usepackage[colorlinks=true
,urlcolor=purple
,anchorcolor=blue
,citecolor=red 
,filecolor=magenta
,linkcolor=blue
,menucolor=blue
,linktocpage=true
,pdfproducer=medialab
,pdfa=true
]{hyperref}
\usepackage[utf8]{inputenc}
\usepackage[english]{babel}
\usepackage{soul}

\newcommand{\mpl}{M_{\rm pl}}

\newcommand{\MeV}{\rm MeV}
\newcommand{\GeV}{\rm GeV}

\newcommand{\es}[2] {\begin{equation} \label{#1} \begin{split} #2 \end{split} \end{equation}}

\begin{document}

\title{
Dark Grand Unification in the Axiverse: Decaying Axion Dark Matter and Spontaneous Baryogenesis
}

\author{Joshua W. Foster}
\email{jwfoster@mit.edu}
\affiliation{Center for Theoretical Physics, Massachusetts Institute of Technology, Cambridge, MA 02139, U.S.A.}

\author{Soubhik Kumar}
\email{soubhik@berkeley.edu}
\affiliation{Berkeley Center for Theoretical Physics, University of California, Berkeley, CA 94720, U.S.A.}
\affiliation{Theoretical Physics Group, Lawrence Berkeley National Laboratory, Berkeley, CA 94720, U.S.A.}

\author{Benjamin R. Safdi}
\email{brsafdi@berkeley.edu}
\affiliation{Berkeley Center for Theoretical Physics, University of California, Berkeley, CA 94720, U.S.A.}
\affiliation{Theoretical Physics Group, Lawrence Berkeley National Laboratory, Berkeley, CA 94720, U.S.A.}

\author{Yotam Soreq}
\email{soreqy@physics.technion.ac.il}
\affiliation{Physics Department, Technion – Israel Institute of Technology, Haifa 3200003, Israel}

\date{\today}
\preprint{MIT-CTP/5458}

\begin{abstract}
The quantum chromodynamics axion with a decay constant near the Grand Unification~(GUT) scale has an ultralight mass near a neV.
We show, however, that axion-like particles with masses near the keV - PeV range with GUT-scale decay constants are also well motivated in that they naturally arise from axiverse theories with dark non-abelian gauge groups.
We demonstrate that the correct dark matter abundance may be achieved by the heavy axions in these models through the misalignment mechanism in combination with a period of early matter domination from the long-lived dark glueballs of the same gauge group. 
Heavy axion dark matter may decay to two photons, yielding mono-energetic electromagnetic signatures that may be detectable by current or next-generation space-based telescopes.
We project the sensitivity of next-generation telescopes including {\it Athena}, AMEGO, and e-ASTROGAM to such decaying axion dark matter.  
If the dark sector contains multiple confining gauge groups, then the observed primordial baryon asymmetry may also be achieved in this scenario through spontaneous baryogenesis.  
We present explicit orbifold constructions where the dark gauge groups unify with the SM at the GUT scale and axions emerge as the fifth components of dark gauge fields with bulk Chern-Simons terms. 
\end{abstract}

\maketitle
\tableofcontents

\section{Introduction}

The quantum chromodynamics~(QCD) axion was originally introduced to explain the strong {\it CP} problem connected to the absence of the neutron electric dipole moment~\cite{Peccei:1977hh,Peccei:1977ur,Weinberg:1977ma,Wilczek:1977pj}.  
The axion is naturally realized as the pseudo-Nambu Goldstone boson of a symmetry, the Peccei-Quinn~(PQ) symmetry, which is spontaneously  broken at a high scale $f_a$.  
The axion $a$ would be exactly massless but for its interactions with QCD through the dimension-5 operator $a G \tilde G / f_a$, where $G$ is the QCD field strength and $\tilde G$ is its dual.  
Below the QCD confinement scale instantons generate a potential for the axion; when the axion minimizes this potential, it dynamically removes the neutron electric dipole moment.  
The axion acquires a mass ${m^{\rm QCD}_a} \approx \frac{\sqrt{m_u m_d}}{m_u+m_d}\frac{m_\pi f_\pi}{f_a}$, with $m_\pi$ ($f_\pi$) the pion mass (decay constant) and $m_u$ ($m_d$) the up quark (down quark) mass.

Coherent fluctuations of the axion field about its minimum may explain the observed dark matter~(DM) abundance~\cite{Preskill:1982cy,Abbott:1982af,Dine:1982ah}.  
If the PQ symmetry is broken after inflation then the correct DM abundance is achieved for $m_a^{\rm QCD} \sim 100$ $\mu$eV~\cite{Gorghetto:2020qws,Buschmann:2021sdq}, while if the PQ symmetry is broken before inflation the final DM abundance depends on the initial misalignment angle, and much lower axion masses may still explain the DM~\cite{Tegmark:2005dy,Hertzberg:2008wr,Co:2016xti,graham2018stochastic,takahashi2018qcd}.

The idea of the `axiverse' naturally emerges in the context of String Theory constructions~\cite{Green:1984sg,Witten:1984dg,Svrcek:2006yi,Arvanitaki:2009fg,Halverson:2019cmy,Conlon:2006tq,Acharya:2010zx,Ringwald:2012cu,Cicoli:2012sz}, whereby there is a large number $N$ of ultralight axion-like particles~(ALPs).
One linear combination of the ALPs couples to QCD and becomes the QCD axion mass eigenstate.  
It is commonly assumed that the rest of the $N-1$ ALPs remain ultralight, with masses much less than the mass of the QCD axion mass eigenstate; the non-zero masses of these ultralight ALPs could arise, {\it e.g.}, from string or gravitational instantons.  
Indeed, in String Theory constructions ALPs arise as the zero modes of higher-dimensional gauge fields compactified on the internal manifold, and the gauge invariance protects the masses of the ALPs from perturbative contributions~\cite{Demirtas:2021gsq}. 
These ultralight ALPs interact with matter and SM gauge fields except for QCD. 
The ALP decay constants may range, roughly, from $f_a \sim 10^{9} - 10^{18}$ GeV, which means that the ALP-matter interactions are heavily suppressed since they arise through higher-dimensional operators suppressed by this scale.  
The upper bound on $f_a$ arises from the theoretical assumption that the decay constant is smaller than the Planck scale, while the lower bound on $f_a$ is determined by stellar cooling and laboratory searches~\cite{Adams:2022pbo}.
There is currently significant effort dedicated to searching for ultralight ALPs and the QCD axion in the laboratory and in astrophysical environments (see~\cite{Adams:2022pbo,Baryakhtar:2022obg,Boddy:2022knd} for recent reviews).  

In this work, we consider the possibility that at least one of the axion\footnote{Throughout the rest of this Article we refer to all ALPs as `axions' for simplicity, with the axion that solves the strong {\it CP} problem distinguished as the QCD axion.} mass eigenstates may be much heavier than the eV scale because of instantons from a dark gauge group. 
In particular, through a coupling $a G_d \tilde G_d / f_a$ to the dark gauge group with gauge field $G_d$, the heavy axion acquires a mass $m_a \sim \Lambda_D^2 / f_a$, with $\Lambda_D$ being the dark confinement scale.  
If the dark gauge group unifies with the Standard Model~(SM) near the Grand Unified Theory~(GUT) scale then we show it is natural to expect $\Lambda_D \sim 10^4 - 10^{10}$ GeV, for low-dimension dark gauge groups such as $SU(2)$ or $SU(3)$.\footnote{See~\cite{Demirtas:2018akl,Cvetic:2020fkd}, however, which claim that in certain F-theory constructions lower confinement scales may be preferred.}
Assuming $f_a$ is also around the GUT scale ($\sim$ $10^{15} - 10^{17}$ GeV) this then implies $m_a\sim {\rm eV}-0.1\,{\rm PeV}$,
though the heavy axion mass could be beyond this range depending on the dark confinement scale and decay constant.  
We show that if the axion is in the keV-MeV mass range it may explain the observed DM abundance.  A crucial ingredient to this story, however, is a period of early matter domination induced by the dark glueballs that dilutes the otherwise over-abundant population of cold axions.

The axion DM may decay today to two photons to give rise to monochromatic $X$-ray and gamma-ray lines.  
Such signatures can be searched for with existing telescopes such as {\it XMM-Newton}, {\it Chandra}, {\it NuSTAR},  INTEGRAL, and COMPTEL~\cite{Boddy:2022knd}. 
Here, we reinterpret existing limits from these telescopes on decaying DM in the context of the heavy axion DM model.  
We further show that future telescopes in the keV-MeV range offer significant discovery potential for heavy axions.  
In particular, we project sensitivity to heavy axion DM from the possible e-ASTROGAM~\cite{e-ASTROGAM:2017pxr} and AMEGO~\cite{Kierans:2020otl} missions in the MeV band that are currently being proposed (see~\cite{Engel:2022bgx} for additional proposals).  
These telescopes will improve the sensitivity to MeV sources by orders of magnitude, and we show that this will provide significant discovery potential for decaying heavy axion DM.  
The recently-launched eROSITA $X$-ray telescope may provide an incremental increase in sensitivity in the $X$-ray band~\cite{eROSITA:2020emt,Dekker:2021bos}, while larger leaps in sensitivity will arise from next-generation $X$-ray telescopes such as {\it Athena}~\cite{Barcons:2015dua, Piro:2021oaa} and THESEUS~\cite{THESEUS:2017qvx} that will probe natural parameter space where we may expect a decaying axion DM model to appear.  
The keV-MeV axions proposed in this work provide strong motivation for pursuing next-generation telescopes across this energy range.  
The observable signature of axion DM decay is similar to that of keV-scale mass sterile neutrino DM~\cite{Drewes:2016upu}, but such models have been increasingly in tension with data~\cite{Perez:2016tcq,Roach:2019ctw,Foster:2021ngm}.  
On the other hand, much of the best-motivated parameter space for heavy axions, as we show, has yet to be covered experimentally.    

As we show, achieving the correct DM abundance without significant fine-tuning of the initial axion misalignment angle requires the reheat temperature from glueball decay to be ${\cal O(\MeV-\GeV)}$. 
With such low values of the reheat temperature, it is interesting to know whether successful baryogenesis can occur.
We demonstrate that in addition to having a keV-MeV mass axion explaining the DM abundance, an even heavier axion state may give rise to the observed baryon asymmetry through spontaneous baryogenesis~\cite{Cohen:1987vi}.  
Spontaneous baryogenesis proceeds through leptogenesis, with the Weinberg operator~\cite{Weinberg:1979sa} providing lepton number violation and the oscillating heavy axion field providing a time-dependent chemical potential for lepton number~\cite{Kusenko:2014uta, Co:2020jtv}. 
Thus a lepton asymmetry can develop in thermal equilibrium. 
This lepton asymmetry is then converted to an initially too large baryon asymmetry through electroweak sphalerons. 
The baryon asymmetry is subsequently diluted to the observed value by the entropy dilution induced from the glueballs of the same dark sector that gives rise to the lighter keV-MeV scale axion DM. 
Therefore, we can explain both the primordial DM and baryon abundances in this scenario with two heavy axions.

As for the case of the axiverse, motivation for considering dark gauge groups arises from String Theory constructions, which
may give rise to non-abelian dark gauge sectors,
including in the hetorotic string, type II string theory models, M-theory, and $F$-theory (see, {\it e.g.},~\cite{Gross:1984dd,Dixon:1985jw,Dixon:1986jc,Ibanez:1986tp,Lebedev:2006kn,Blaszczyk:2009in,Braun:2005ux,Bouchard:2005ag,Anderson:2012yf,Cvetic:2004ui,Gmeiner:2005vz,Blumenhagen:2008zz,Acharya:1998pm,Halverson:2015vta,Grassi:2014zxa,Halverson:2015jua,Taylor:2015ppa,Acharya:2017szw}).
In particular, at energy scales well below the GUT scale, the gauge group may be written as $G_{\rm SM} \times G_{\rm dark}$, where $G_{\rm SM}$ is the SM gauge group and $G_{\rm dark}$ is the dark gauge group.  
No SM matter is charged under $G_{\rm dark}$.
In this work we show, through explicit constructions based on, {\it e.g.}, an $SU(8)$ group, that such dark gauge sectors can also arise in orbifold GUT models where $G_{\rm SM}$ and $G_{\rm dark}$ unify into a single non-abelian gauge group.
Note that while dark glueballs may themselves be the DM,
as discussed first in~\cite{Carlson:1992fn} and further elaborated upon in {\it e.g.}~\cite{Halverson:2016nfq,Faraggi:2000pv,Feng:2011ik,Boddy:2014yra,Soni:2016gzf,Kribs:2016cew,Acharya:2017szw,Yamanaka:2019aeq,Halverson:2018olu,Halverson:2018vbo,Acharya:2017kfi,Soni:2017nlm,Cohen:2016uyg,Asadi:2021pwo},
in this work we assume the glueballs decay before big bang nucleosynthesis~(BBN).  Scalar (moduli) DM may also arise in String Theory constructions with similar phenomenology to the heavy axions discussed in this work~\cite{Kusenko:2012ch}.
For previous discussions of the unification of dark gauge groups with the SM see, {\it e.g.}, Ref.~\cite{Gherghetta:2016fhp, Gaillard:2018xgk, Murgui:2021eqf}.

The remainder of this Article is organized as follows. 
In Sec.~\ref{sec:axiverse} we discuss the field theory of multiple axions connected to non-abelian, confining dark sectors. We describe the axion masses, decay constants, and couplings to matter that would naturally arise in the presence of such dark sectors. 
In Sec.~\ref{sec:DM} we describe the cosmology in the presence of the associated glueballs of the confining dark sector. 
We show that such glueballs give rise to an early matter dominated era and naturally avoid the DM overclosure problem, making the heavy axions a suitable DM candidate. 
The resulting axion DM can decay into a pair of photons, and we show that the existing and proposed $X$-ray and gamma-ray missions are capable of probing much of the motivated parameter space.
In Sec.~\ref{sec:bary} we first show that an even heavier axion can lead to successful baryogenesis through the mechanism of spontaneous baryogenesis.
Subsequently, we discuss a scenario in which the presence of two heavy axions can explain both the observed DM and baryon abundances through their connected cosmological evolution.
To give an example of how such confining dark sectors might arise, in Sec.~\ref{sec:model} we construct an extra-dimensional orbifold GUT model, describing a breaking pattern $SU(8)\rightarrow SU(3)_D \times G_{\rm SM}$.  
The dark axion naturally emerges in this scenario from a higher-dimensional gauge field. We conclude in Sec.~\ref{sec:conclu}.

\section{keV - PeV axions from confining dark sectors}
\label{sec:axiverse}

In this section we motivate keV-PeV scale axions from confining dark sectors.
We claim that such heavy axion states arise generically in axiverse models with dark gauge groups.  
In particular, we assume that well below the GUT scale, the gauge group of nature may be written as $G_{\rm SM} \times G_{\rm dark}$.
For simplicity we assume that the dark sector has no light matter content. Importantly, all interactions between the dark sector and the visible sector occur through higher-dimensional operators.  
We consider a scenario where $G_{\rm SM}$ and $G_{\rm dark}$ are unified at some high scale $M_{\rm GUT} \sim 10^{15}-10^{17}$~GeV, whereas they still interact through an intermediate scale $\Lambda \lesssim M_{\rm GUT}$.
Explicit, example constructions of $G_{\rm SM}$ and $G_{\rm dark}$ unification in an extra dimensional framework are given in Sec.~\ref{sec:model}. 

\subsection{Field theory considerations in the axiverse with confining dark sectors}

The confinement scale of the dark sector $\Lambda_D$ is related to the ultraviolet~(UV) coupling $\alpha_{\rm UV}$ at the energy scale $\Lambda_{\rm UV}$ via the relation 
\es{eq:dark_confinement}{
    \Lambda_D 
=   \Lambda_{\rm UV} \, {\rm exp}
    \left({- \frac{2 \pi}{  3 T_{G_{\rm dark}}\alpha_{\rm UV}}}  \right)\,,
}
where $T_{G_{\rm dark}}$ is the dual Coxeter number, which is $N$ for $G_{\rm dark} = SU(N)$, and assuming that the dark sector is ${\mathcal N} = 1$ supersymmetric.
If the dark gauge group is not supersymmetric then we may use the one-loop $\beta$-function to estimate the dark confinement scale, which leads to an analogous expression to~\eqref{eq:dark_confinement} but with $T_{G_{\rm dark}} \to {11 } C_2 /9$, with $C_2$ the quadratic Casimir ($C_2 = N$ for $SU(N)$).  

We assume no light matter charged under $G_{\rm dark}$.
Let us suppose that the dark sector unifies with the visible sector at $\Lambda_{\rm GUT} \approx 10^{16}$ GeV, with $\alpha_{\rm UV} \approx 1/24$, as motivated by supersymmetric grand unification~\cite{Mohapatra:1997sp}.  
Then, taking $G_{\rm dark} = SU(2)$ we find $\Lambda_D \approx 10^5$ GeV ($\Lambda_D \approx 10^7$ GeV) for the supersymmetric (non-supersymmetric) theory; if instead $G_{\rm dark} = SU(3)$ then the dark confinement scale rises to $5 \times 10^8$ GeV and $10^{10}$ GeV for the supersymmetric and non-supersymmetric theories, respectively.  
Matter content in the dark sector may further lower the confinement scales.  
Moreover, the assumed value $\alpha_{\rm UV}  \approx 1/24$ is suggestive but may deviate in any particular GUT model, which broadens the possible confinement scales.  
Thus, for most of this work we remain somewhat agnostic as to the scale $\Lambda_D$, though high confinement scales $\sim 10^5 - 10^{10}$ GeV appear to be natural expectations.   

We assume that there are $N$ axions $a_i$, with $i = 1, \cdots, N$, that have ultralight bare masses (much lighter than the QCD axion mass).  
The axions will acquire non-trivial potentials through their couplings to $G_{\rm dark}$ and to the visible $SU(3)_{c}$.  
In principle, $G_{\rm dark}$  may have multiple confining sub-sectors. 
For the moment, however, we take $G_{\rm dark}$ to be a simple Lie group, whose confinement scale is assumed to be much larger than $\Lambda_{\rm QCD}$. (Later, we consider the scenario where $G_{\rm dark}$ is the product of two simple Lie groups, one of which gives rise to the DM axion and the other produces the heavy axion that leads to baryogenesis.)  
We denote the field strength as $G_{\rm d, \mu\nu}^a$, with $a$ a dark color index.  
Then, the relevant terms in the Lagrangian are
\es{}{
    {\mathcal L} 
=   \sum_i {\alpha_d \over 8 \pi}{c_{\rm d}^i a_i \over f_a} 
    G_{{\rm d} \, {\mu \nu}}^a \tilde G_{{\rm d}\,a}^{\mu \nu}  \,,
}
where $f_a$ is the decay constant giving the scale of the ultraviolet completion to the axion sector, $\alpha_d$ is the dark gauge coupling, and the $c_{\rm d}^i$ 
are dimensionless coefficients that describe the magnitude of the coupling of each axion to the dark gauge group.  
Effectively we treat the axions $a_i$ to have decay constants $f_a / |c_d^i|$, but we chose to factor out the common scale $f_a$; in particular, we assume that in the UV completion each axion field $a_i$ is periodic with period $2 \pi f_a / |c_d^i|$.
At energies well below the dark confinement scale, instantons in the dark sector generate a potential for the axions, which for small displacements is of the form
\es{eq:pot}{
    V \approx 
    &\Lambda_{D}^4 \left( \sum_i  {c_{\rm d}^i a_i \over f_a} + \bar \theta_D \right)^2 \,,
}
where ${\bar \theta_D}$ 
is the CP-violating theta-angle of the dark sector.
The canonically normalized axion mass eigenstate is given by
\es{}{
a_D = {\sum_i c_{\rm d}^i a_i \over \sqrt{\sum_i (c_{\rm d}^i)^2 }} \,,
}
and the axion mass is 
\es{eq:axion_mass}{
m_a \approx {\Lambda_D^2 \over \tilde f_a} \approx 1 \, \, {\rm MeV} \left( {\Lambda_D \over 10^6 \, \, {\rm GeV}} \right)^2 \left( {10^{15} \, \, {\rm GeV} \over \tilde f_a}\right) \,.
}
We define $\tilde f_a \equiv f_a / \sqrt{\sum_i (c_{\rm d}^i)^2 }$, such that the axion coupling to the dark gauge group is
\es{}{
{\mathcal L} =  {\alpha_d \over 8 \pi}{a_D \over \tilde f_a} G_{{\rm d} \, {\mu \nu}}^a \tilde G_{{\rm d}\,a}^{\mu \nu}  \,.
}
If we assume that all of the $c_{\rm d}^i$ are order unity, then $\tilde f_a \sim f_a / \sqrt{N}$.  
The axion $a_D$ has domain wall number $N$ 
in this construction
with respect to the dark gauge group; that is, $a_D$ is periodic with period $N \times 2 \pi \tilde f_a$,
but the dark-gauge-group-induced potential is periodic with period $2 \pi \tilde f_a$.

Let us now consider the couplings of the axion $a_D$ to other gauge sectors.  
In particular, we assume that the ensemble of axions, $a_i$, interact with a gauge group specified by field strength $G_{\mu\nu}$ and coupling strength $\alpha$ by the terms
\es{eq:other_sector}{
    {\mathcal L} 
=   \sum_i {\alpha \over 8 \pi} {d^i a_i \over f_a} G_{\mu \nu} \tilde G^{\mu \nu} 
=   {\alpha \over 8 \pi} {d_D a_D \over \tilde f_a} G_{\mu \nu} \tilde G^{\mu \nu} 
    + \cdots  \,,
}
where the $d^i$ are dimensionless constants.  
Note that this gauge group may represent an additional confining dark gauge group, the visible QCD sector, or $U(1)_{\rm EM}$; the point we make about this coupling is generic assuming that the confinement scale for $G_{\mu \nu}$, if it confines, is much lower than $\Lambda_D$.
In the second equality in~\eqref{eq:other_sector} we have isolated the interaction of $a_D$ and left off the other $N-1$ axion states. 
The dimensionless coupling $d_D$ can be written as
\es{}{
d_D = {\sum_i c_d^i d^i \over {\sum_i (c_d^i)^2}} \,.
}

Under the assumption $\langle c_d^i \rangle = \langle d^i \rangle = \langle c_d^i d^i \rangle = 0$ and $\langle (c_d^i)^2 \rangle = \langle (d^i)^2 \rangle = 1$, with brackets denoting correlations over statistical realizations of the couplings, then $\langle d_D^2 \rangle \approx 2 / N$.
This is important because it suggests that in an axiverse with $N$ axions the couplings of massive axion states to gauge groups with lower confinement scales will be suppressed by $\sim$ $1/\sqrt{N}$.  
Of course, the exact suppression depends upon the distributions of axion couplings to the various gauge groups, but generically we may expect the couplings of the massive axion state to the other gauge groups to be suppressed.  (See~\cite{Halverson:2019kna} for a similar observation in the context of axion reheating through couplings to gauge sectors in F-theory.)

As an aside from the heavy axion discussion, consider the IR
coupling of the QCD axion to electromagnetism, at energy scales below the QCD confinement scale in the context of the axiverse:
\es{}{
    {\mathcal L} 
=  C_{a\gamma\gamma} {\alpha_{\rm EM}  \over 8 \pi}     
     {a_{\rm QCD} \over \tilde f_a} \tilde F_{\mu \nu} F^{\mu \nu} \,,
}
where we have identified $\tilde f_a$ with the decay constant of the QCD axion mass eigenstate, which is generically a factor of $1 / \sqrt{N}$ smaller than the decay constants of the $a_i$ axion states, assuming the appropriate $c_d^i$ coefficients are order unity and uncorrelated.  
Note that for this particular discussion the presence of a possible dark gauge sector does not play an important role. 
The IR coefficient $C_{a\gamma\gamma} = C_{a\gamma\gamma}^{\rm UV} + C_{a\gamma\gamma}^{\rm QCD}$ has an ultraviolet contribution $C_{a\gamma\gamma}^{\rm UV}$ and a contribution from mixing of the neutral pion, 
$C_{a\gamma\gamma}^{\rm QCD} \approx -1.92(4)$~\cite{diCortona:2015ldu}.  
The UV
contribution is typically written as $C_{a\gamma\gamma}^{\rm UV} = E_{\rm QCD}/ N_{\rm QCD}$, where $E_{\rm QCD}$\,($N_{\rm QCD}$) is the electromagnetic\,(QCD) anomaly coefficient.  
The argument above suggests that in an axiverse with $N$ nearly degenerate (in decay constant) axions, we expect the QCD mass eigenstate to have $N_{\rm QCD} \sim N$ while $E_{\rm QCD} \sim \sqrt{N}$, and thus $E_{\rm QCD} / N_{\rm QCD} \sim 1 / \sqrt{N}$.  This then implies that the infrared observer should measure $C_{a\gamma\gamma} \approx C_{a \gamma\gamma}^{\rm QCD}$, which is the expectation for the KSVZ field theory axion model~\cite{Kim:1979if,Shifman:1979if}.  
In contrast, in models where the QCD axion couples to the SM in a way compatible with Grand Unification we expect $E_{\rm QCD} / N_{\rm QCD} = 8/3$ (see, {\it e.g.},~\cite{DiLuzio:2020wdo}), leading to the DFSZ-type expectation $C_{a\gamma\gamma} \approx 0.75$~\cite{Dine:1981rt,Zhitnitsky:1980tq}.  
Note that the axiverse scenario could still lead to the DFSZ-type $C_{a\gamma\gamma}$ if all of the axions share the same GUT-compatible coupling to the SM gauge groups, as that would violate our assumption that the $c^i$ are uncorrelated.  
For a recent discussion along these lines see~\cite{Agrawal:2022lsp}.  

Moreover, we note that the QCD axion decay constant, as defined through the coupling of the axion to QCD, is reduced by a factor $\sim$$ \sqrt{N}$ from the naive expectation in the axiverse, assuming uncorrelated coupling coefficients. (The decay constant would be reduced by $\sim$$N$ if the couplings are correlated.)
This has important implications for axion laboratory experiments such as ABRACADABRA and DM Radio~\cite{Ouellet:2018beu,Salemi:2021gck,Brouwer:2022bwo,DMRadio:2022pkf}, as it suggests that decay constants as low as, {\it e.g.}, $\sim 10^{13} - 10^{14}$\,GeV could be directly connected with GUT models in the context of the axiverse with a large $N$ number of axions. 

Returning to the heavy axion story, we note that the same logic applied above to the QCD axion also suggest that heavy axion axion-photon coupling coefficients $C_{a\gamma\gamma} \sim 1 /\sqrt{N}$ might be expected in the axiverse, as the heavy axions have only UV contributions to the electromagnetic coupling.  
For example, $C_{a\gamma\gamma} \sim 0.1$ could be expected for $N \sim 10^2$ axions. 

In addition to the axion-photon coupling we also consider the axion-electron coupling, which for an axion $a$ is 
\es{}{
{\mathcal L} = C_{aee}{ \partial_\mu a \over 2 \,f_a} \bar e \gamma^\mu \gamma_5 e \,,
}
where $C_{aee}$ is the dimensionless coupling coefficient and $e$ is the electron field. 
Depending on the UV completion this coefficient may be zero or non-zero in the UV, though given an axion-photon coupling it is generated at one-loop under the renormalization group. 
We use this operator when considering axion decays to electron-positron pairs, where kinematically allowed.

\section{Axion cosmology with early matter domination from dark glueballs}
\label{sec:DM}

In this section we discuss heavy axion cosmology and show, in particular, that the correct DM abundance may naturally arise if there is a period of early matter domination.  
The early matter domination, ending with a low reheat temperature $T_{\text{RH}}$, can naturally arise in the context of the heavy axion theory, with no additional ingredients beyond the heavy axion and its associated dark gauge group, because of the dark glueballs.

\subsection{Signatures of heavy axion dark matter}

Let us first suppose that there is a heavy axion in the keV-MeV mass range, whose mass is generated from a confining dark sector as described previously,  that makes up the observed DM.  
If the axion mass is less than twice the electron mass then the only kinematically-allowed option for the axion to decay is into two photons.  
(Note that heavy axion decays to lighter axions will generically be suppressed relative to axion decays to two photons and to electron-positron pairs.)
The decay rate of the axion to two photons is given by 
\es{eq.lifetime}{
\tau_{a\to \gamma\gamma}=\frac{256\pi^3}{\alpha^2 C_{a\gamma\gamma}^2}\frac{f_a^2}{m_a^3}\approx \, &9.6\times 10^{27}~\text{s}\left(\frac{0.1}{C_{a\gamma\gamma}}\right)^2 
\left(\frac{0.1~\text{MeV}}{m_a}\right)^3 \left(\frac{f_a}{10^{15}~\text{GeV}}\right)^2 \,.
}
Above, and in the remainder of this Article, we depart from the notation in the previous section for simplicity and take $\tilde f_a \to f_a$ to be the axion decay constant such that $m_a \approx \Lambda_D^2 / f_a$ (see~\eqref{eq:axion_mass}).  
Interestingly, while much longer than the age of the Universe, DM lifetimes on the order of those in~\eqref{eq.lifetime} are at the edge of sensitivity of present-day $X$-ray and gamma-ray telescopes, as we discuss later in this Article. 

When $m_a > 2 m_e$, with $m_e$ the electron mass, the axion may also decay to $e^+ e^-$ pairs, with partial lifetime (see, {\it e.g.},~\cite{Bauer:2019gfk}) 
\es{}{
\tau_{a \to e^+e^-} &= {8 \pi f_a^2 \over m_a m_e^2} {1 \over C_{aee}^2} \left[ 1 - 4 {m_e^2 \over m_a^2}\right]^{-1/2}\\
&\approx {4 \times 10^{18}} \, \, {\rm s} \left({0.1 \over C_{aee}}\right)^2 \left( {2 \, \, {\rm MeV} \over m_a} \right) 
\left(\frac{f_a}{10^{15}~\text{GeV}}\right)^2
\,.
}
Thus, the axion to $e^+e^-$ pair decay channel may dominate the total lifetime.  
In fact, this may be true even if $C_{aee}$ vanishes in the UV and is only generated under the renormalization group. 
The IR value of $C_{aee}$ in that case depends on the relative coupling of the axion to $U(1)_Y$ versus $SU(2)_L$, but one generically expects $|C_{aee} / C_{a\gamma\gamma}| \sim 10^{-4} - 10^{-3}$~\cite{Srednicki:1985xd,Chang:1993gm,Dessert:2021bkv}.  
Thus, depending on $|C_{aee} / C_{a\gamma\gamma}|$, the total lifetime for $m_a \gtrsim 2 m_e$ could be dominated by the decay to $e^+e^-$ pairs.  
The total lifetime must be sufficiently long compared to the age of the Universe for the axion to be a DM candidate.  
This requirement itself limits the $f_a$ that may be realized for tree-level $C_{aee}$ and $m_a \sim {\mathcal O}(1)$ MeV.  
However, the constraints on $\tau_{a\to\gamma\gamma}$ are much stronger than those on $\tau_{a\to e^+e^-}$. 
For example, for $m_a\sim2\,\MeV$, the lower bound on the axion decay rate to photons is $\tau_{a\to\gamma\gamma} \gtrsim {\rm few} \times 10^{27}\,$s, while for electrons it is $\tau_{a\to e^+e^-} \gtrsim 10^{24}$\,s~\cite{Liu:2020wqz}.
Thus, we conclude that for $m_a > 2 m_e$, DM decays to $e^+e^-$ pairs generically rule out axion DM with tree-level couplings to electrons, for $f_a$ all the way up towards the Planck scale, while in the case of loop-induced axion-electron couplings the probes using decays to two photons are more powerful. 
For this reason, throughout the rest of this Article we assume that for $m_a > 2 m_e$ the axion-electron coupling is loop induced, so that we may focus solely on the decay channel to $\gamma\gamma$.

\subsection{Cosmology of heavy axion dark matter with and without early matter domination}

A central impediment, however, to the possibility of keV-MeV scale axion DM is that assuming the standard cosmology, DM is overproduced by orders of magnitude.  
From the misalignment mechanism, assuming the axion field starts with a constant initial field value $a_D(t=0) = \theta_i f_a$ and its mass $m_a$ is temperature independent, the DM abundance is determined to be
\es{eq.relicab_std}{
    \Omega_a h^2\rvert_{\rm RD}
    \approx 
    0.12\left(
    \frac{\theta_if_a}{2\times10^{13}~\text{GeV}}
    \right)^2 
    \left(\frac{m_a}{1~\mu\text{eV}}\right)^{\frac{1}{2}} 
    \left(\frac{90}{g_{*}(T_{\rm osc})}\right)^{\frac{1}{4}} \,,
}
in the limit $|\theta_i| \lesssim 1$ where anharmonicities of the axion potential may be ignored.
Here $g_{*}(T_{\rm osc})$ is the effective number of degrees of freedom in the radiation bath when the axion starts to oscillate at $m_a=q_0 H(T_{\rm osc})$, with $q_0 \approx 1.6$ (see {\it e.g.}~\cite{Blinov:2019rhb}). 
(Here we ignore possible temperature dependence of the axion mass, though we will return to this possibility later in this Article.) 
Given that cosmic microwave background~(CMB) measurements indicate $\Omega_a h^2 \approx 0.12$~\cite{Planck:2018vyg}, masses $m_a \sim$~keV-MeV appear heavily disfavored, unless the initial misalignment angle is severely tuned.  
One possibility is that the tuning could appear for anthropic reasons, as has been discussed, {\it e.g.}, for the QCD axion with GUT-scale decay constant~\cite{Tegmark:2005dy}. 
However, in this Article we explore dynamical mechanisms that may create the correct DM abundance without requiring anthropic tuning of $\theta_i$.  

A key point of this work is that we may naturally match the observed abundance of DM for such massive axions by assuming that the Universe went through a period of early matter domination.  
We will later show that the early matter domination may arise from the dark glueballs.   
The DM abundance from the misalignment mechanism may generically be written as~\cite{Blinov:2019rhb} 
\es{eq:gen_ev}{
\Omega_a = {1 \over 2} {m_a^2 f_a^2 \theta_i^2 \over \rho_c} \left( {R_{\rm osc} \over R_{\rm RH}} \right)^3 \left( {T_0 \over T_{\rm RH}} \right)^3 {g_{*S}(T_0) \over g_{*S}(T_{\rm RH})} \,,
}
where $T_{\rm RH}$ is the reheat temperature after early matter domination, assuming instantaneous reheating, $T_0$ is the temperature today, $\rho_c$ is the critical density, and $R_{\rm osc}$ ($R_{\rm RH}$) is the scale factor at $T_{\rm osc}$ ($T_{\rm RH}$).  
The evolution is assumed to be adiabatic below $T_{\rm RH}$, and we assume for now that $m_a$ is temperature independent.  
The scale factor ratio appearing in~\eqref{eq:gen_ev} may be simplified by using the equation of state $H^2 \propto R^{-3(w+1)}$, with $w = 0$ ($w = 1/3$) for matter (radiation) domination.  
Assuming a standard radiation dominated cosmology in~\eqref{eq:gen_ev}, in which case $T_{\rm RH}$ is any intermediate reference temperature, then leads to the result quoted in~\eqref{eq.relicab_std}.  
On the other hand, if we suppose that the axion starts to oscillate during a period of early matter domination, with instantaneous reheating at $T_{\rm RH}$, then the dependence of $\Omega_a$ on $m_a$ and $g_{*S}(T_{\rm RH})$ cancels, leading to the result~\cite{Blinov:2019rhb}
\begin{align}
\label{eq.relicab}
\left.\Omega_a h^2\right|_{\rm EMD}\approx0.12\left(\frac{\theta_if_a}{10^{15}~\text{GeV}}\right)^2\left(\frac{T_{\text{RH}}}{10~\text{MeV}}\right).
\end{align}
Note that if the axion starts to oscillate during radiation domination, with the Universe subsequently going through a period of early matter domination, the expression for $\Omega_a h^2$ in~\eqref{eq.relicab} is enhanced by the ratio $(T_{\rm osc} / T_{\rm EMR})$, where $T_{\rm EMR}$ is the temperature of matter-radiation equality at the beginning of the early matter domination epoch.
Thus, we see heavy axions can indeed constitute all of DM for GUT-scale $f_a$ without fine tuning in the initial misalignment $\theta_i$, provided $T_{\text{RH}}$ is sufficiently low.  
In the next subsection we show that such a period of early matter domination, with low reheating temperature, may naturally arise from glueballs in the dark sector.
Note that successful BBN requires $T_{\rm RH}>4$~MeV~\cite{Kawasaki:2000en, Hannestad:2004px}; {\it i.e.}, the glueballs must have decayed to give rise to a radiation dominated cosmology below this temperature. 
Therefore this determines the lower limit of $T_{\rm RH}$ in the subsequent analyses.

\subsection{Early matter domination from dark glueballs}

In the absence of any fermionic states in the dark sectors, the glueballs arising from confinement of $G_{\rm dark}$ would typically be long-lived. 
They can still decay into SM states through higher-dimensional couplings to the SM Higgs, which we parameterize by 
\es{eq:glueball_dim6}{
{\mathcal L} \supset  &{c_6 \alpha_D \over 4 \pi} G^a_{d \, \mu \nu} {G}_{d \, a}^{ \mu \nu} {H^\dagger H \over \Lambda^2} 
+{\tilde c_6 \alpha_D \over 4 \pi} G^a_{d \, \mu \nu} {\tilde G}_{d \, a}^{ \mu \nu} {H^\dagger H \over \Lambda^2} \,.
}
Here $\Lambda$ is a generically high scale and can be of the order of the GUT-scale or  lower. 
The dimensionless coefficient $c_6$ and $\tilde c_6$ will generically both be of order unity if the UV theory has CP violation and if the dimension-6 operators are generated at one-loop by couplings to heavy particles that interact with the SM Higgs and are charged under the dark gauge group. 
We provide an explicit construction of this operator along these lines in Sec.~\ref{sec:model} in this Article.

Let us assume that the lightest glueball,
which is the $J^{PC} = 0^{++}$ state, has mass $m_{0^+}$.  
We focus on the $c_6$ operator since it is the relevant one for the decay of the $0^{++}$ glueball in the limit of vanishing $\theta$-angle, which is accomplished by the dark axion.\footnote{A residual, oscillating $\theta$ angle may be present from the oscillating axion field about its minimum, but this possibility does not affect the arguments below, as it would simply introduce a small, time-dependent mixing between the CP even and odd glueball states.} Furthermore, in a CP-violating theory, the heavier glueball states would be even more unstable, and therefore we consider only the $0^{++}$ glueball from here on.\footnote{Note that some of the higher-spin glueballs, such as the $1^{+-}$ state, may require higher-dimensional operators to decay; however, the relic DM abundance of these states is subdominant in the parameter space we consider~\cite{Forestell:2016qhc}. } It is convenient to define the matrix element $ \langle 0 | G^a_{d \, \mu \nu} {G}_{d \, a}^{ \mu \nu} | 0^{++} \rangle \equiv 2 F_{0^+}$ and the dimensionless constant $f_{0^+}$ through the relation $4 \pi \alpha_D F_{0^+} = f_{0^+} m_{0^+}^3$, in order to factor out renormalization group and scale dependence.  
The quantity $f_{0^+}$ has been computed in lattice QCD for pure $SU(3)$ gauge theory to be $f_{0^+} \approx 3.06$, with the dependence on the number of colors $N_c$ expected to be minor for $N_c \sim 3$~\cite{Chen:2005mg}.  
We may then parameterize the glueball decay rate, in the limit $m_{0^+} \gg m_h$, with $m_h$ the Higgs mass, as~\cite{Juknevich:2009gg}
\es{eq:glueball_decay}{
\Gamma_{0^{++} \to {\rm SM}} \approx &~9\, \times 10^{-2} \, \, {\rm s}^{-1} c_6^2 \left( {m_{0^+} \over 10^7 \, \, {\rm GeV} } \right)^5 \left( { 10^{14} \, \, {\rm GeV}} \over \Lambda \right)^4  \,.
}
Note that the glueball decays to pairs of SM Higgs bosons, pairs of $Z$ bosons, and $W^+ W^-$ with relative rates $1:1:2$, respectively.

The ratio $x \equiv m_{0^+} / \sqrt{m_a f_a}$ is expected to be order unity and is independent of the dark confinement scale $\Lambda_D$, though it may have minor dependence on the number of dark colors.  
This ratio is important for determining the cosmological history for the simple reason that the glueball decay rate in~\eqref{eq:glueball_decay} depends on $m_{0^+}$ to the fifth power, so factors order unity in the relation between the confinement scale and the glueball mass become amplified.  
At a more precise level, the axion mass is related to the topological susceptibility $\chi_t$ through the relation $m_a^2 = \chi_t / f_a^2$.  
The topological susceptibility has been computed in lattice QCD for $SU(2)$ and larger $N_c$ gauge theories~\cite{Athenodorou:2021qvs}. 
The glueball mass spectrum has also been computed in lattice QCD~\cite{Bonanno:2022yjr}. 
Combining these results we estimate $x \approx 7.92$ ($x \approx 8.35$)  for $SU(2)$ ($SU(N_c\rightarrow\infty)$) gauge theory.  
Given that the dependence on $N_c$ is minor, in the following calculations we simply take $x = 8$.  

We assume that after inflation both the dark sector and the visible sector are reheated; we denote the ratio of entropy densities between the two sectors as $B \equiv S_D / S_{\rm vis}$, with $S_D$ ($S_{\rm vis}$) the dark (visible) sector entropy density.  
In this section we work in the limit  $B \gg 1$, though our results are not sensitive to $B$ so long as $B \gtrsim 1$ (though $B \ll 1$ is a qualitatively different regime). 
We take the dark confinement phase transition to take place at the critical temperature $T_c$.  
For $T > T_c$, with $T$ the temperature in the dark sector, the dark-sector energy density, which is the dominant energy density in the Universe by assumption, redshifts as radiation.  

We work in the approximation where at $T = T_c$ the dark gluons are instantaneously converted to glueballs, which then redshift like non-relativistic matter.  This approximation is justified because the duration of the phase transition is expected to be much less than a Hubble time with a small degree of supercooling (see, {\it e.g.},~\cite{Halverson:2020xpg}). 
See, {\it e.g.},~\cite{Carlson:1992fn, Forestell:2016qhc,Soni:2016gzf} 
for more careful computations of ${\cal O}(1)$ corrections to this approximation accounting for glueball freezeout and number changing processes.  
Thus, for $T < T_c$ the Universe is matter dominated.  
Matter domination ends when the $0^{++}$ glueball decays to SM final states, leading, in the instantaneous decay approximation, to a visible-sector reheat temperature $T_{\rm RH}$ determined by:  
\begin{align}
\frac{\pi^2}{30}g_{*}(T_{\rm RH})T_{\rm RH}^4 = {4 \over 3} \mpl^2 \Gamma_{0^{++} \to {\rm SM}}^2 \,,
\end{align}
with $g_*(T_{\rm RH})$ the degrees of freedom in the visible sector at $T_{\rm RH}$.  
To derive the right hand side above, we assume that the glueballs instantaneously decay at $t = \Gamma_{0^{++} \to {\rm SM}}^{-1}$, and we use the fact that glueball energy density determines the Hubble parameter during the period of early matter domination.  
Note that this implies a reheating temperature
\es{eq:TRH_glue}{
T_{\rm RH} \approx &~5 \, \, {\rm MeV} \left( {10.8 \over g_*(T_{\rm RH}) }\right)^{1/4} c_6 
\left( {m_{0^+} \over 3 \times 10^7 \, \, {\rm GeV}} \right)^{5/2} \left({10^{14} \, \, {\rm GeV} \over \Lambda} \right)^2 \,.
}
Referring back to~\eqref{eq.relicab}, we see that this reheating temperature is sufficiently low such that we may have $\Lambda$ and $f_a$ in the range $\sim$ $10^{14}$-$10^{16}\,$GeV and produce axions that make up the observed DM, without the need for (much) fine-tuning of the initial misalignment angle. 

However, in the above discussion we crucially make the assumption that the heavy axion starts oscillating {\it during} matter domination.  
Let us revisit this assumption to see under what conditions it holds.  
The ratio of the critical temperature to the topological susceptibility may be computed using lattice QCD results~\cite{Lucini:2012wq} as
\es{}{
{T_c \over \sqrt{m_a f_a}} \approx 1.6 - {0.8 \over N_c^2}
\label{eq:suscep}\,,
}
for $N_c$ dark colors.  
Recall that the axion field begins to oscillate at the temperature $T_{\rm osc}$ where $m_a = q_0 H(T_{\rm osc})$; for $T > T_c$, $H \approx {\pi \over \sqrt{90}} \sqrt{g_*} T^2 / \mpl$, where $g_* = 2(N_c^2 -1)$ is the number of degrees of freedom in dark gluons.  Thus, we see that the dark axion begins to oscillate at $T<T_c$ if $f_a \gtrsim 9.4 \times 10^{17}$ GeV ($f_a \gtrsim 1.2 \times 10^{18} / N_c$ GeV) for $SU(2)$ ($SU(N_c)$ with $N_c \gg 1$), in which case the axion begins to oscillate during matter domination.

Let us consider a benchmark scenario for which the axion beings to oscillate during matter domination.  
We take $N_c = 2$ and saturate $f_a = 9.4 \times 10^{17}$\,GeV, such that $T_{\rm osc} = T_c$. 
Then, requiring a reheat temperature of $T_{\rm RH} \approx 5$\,MeV means that the correct DM density is only achieved if the initial misalignment angle is tuned such that $|\theta_i| \approx 1.5 \times 10^{-3}$ (see~\eqref{eq.relicab}).  
The tuning may be partially alleviated by having the axion begin its oscillation before $T_c$, as we discuss now.

For $T_{\rm osc} > T_c$, we need to account for the temperature dependent axion mass to determine $T_{\rm osc}$, since $m_a(T) \to 0$ for $T$ much larger than $T_c$, while $m_a(T)$ asymptotes to its zero temperature value $m_a$ at $T \approx T_c$.  For $T/T_c \gg 1$ we may reliably use the dilute instanton gas approximation (DIGA) to calculate~\cite{Callan:1977gz}:
\begin{align}
    m_a(T) \approx
    \begin{cases}
		m_a \left(\frac{T_c}{T}\right)^b,~ T>T_c\\
		m_a,~ T\leq T_c \,,
	\end{cases}
\end{align}
where $b = 11 N_c/6 - 2$ for pure $SU(N_c)$.  We then solve for $T_{\rm osc}$ by setting $m_a(T_{\rm osc}) = q_T H(T_{\rm osc})$, with $q_T = (4/5)(2+b)$~\cite{Blinov:2019rhb}, which yields the ratio 
\es{eq:Tosc_over_Tc}{
{T_{\rm osc} \over T_c }&\approx \left( {3 \sqrt{10} \over q_T \pi \sqrt{g_*} \left(1.6 - {0.8 \over N_c^2} \right)^2} {\mpl \over f_a} \right)^{ 1 \over 2 + b} 
\approx
\begin{cases}
	5.5 \left( {10^{15} \, \, {\rm GeV} \over f_a}\right)^{0.27},~N_c = 2\\
2.6 \left( {10^{15} \, \, {\rm GeV} \over f_a}\right)^{0.18},~N_c = 3
\end{cases}
\,.
}
The DM abundance is computed starting from the number density of axions present at $T_{\rm osc}$: $n_a(T_{\rm osc}) = {1 \over 2} m_a(T_{\rm osc}) f_a^2 \theta_i^2$.  The number density is diluted over time, such that at reheating the energy density in axions is $\rho_a(T_{\rm RH}) = {1 \over 2} m_a(T_{\rm osc}) m_a f_a^2 \theta_i^2 \left( R_{\rm osc} / R_{\rm RH}\right)^3$. 
The scale-factor ratio is given by
\begin{align}
	\left(\frac{R_{\rm osc}}{R_{\rm RH}}\right)^3 = \left(\frac{T_c}{T_{\rm osc}}\right)^3\left(\frac{H_{\rm RH}}{H_c}\right)^2 \,,
\end{align}
where $H_c$ is the Hubble rate at $T_c$.  
Further evolving $\rho_a$ down to today and comparing to $\rho_c$ yields the result 
\es{eq:mod_relic_ab}{
\Omega_a h^2 \approx 0.12 \, \theta_i^2 \begin{cases}
	\left(\frac{f_a}{10^{13}~\text{GeV}}\right)^{1.27 }\left(\frac{T_{\rm RH}}{10~\text{MeV}}\right) \,,~N_c = 2 \\
	\left(\frac{f_a}{4.3\cdot 10^{12}~\text{GeV}}\right)^{1.18}\left(\frac{T_{\rm RH}}{10~\text{MeV}}\right) \,,~N_c = 3.
\end{cases} 
}
For $N_c=2$ this implies that for an order one initial misalignment angle and a reheat temperature $T_{\rm RH} \approx 10$\,MeV the correct DM abundance is obtained for $f_a \approx 10^{13}$\,GeV. 
If we require $f_a \approx 10^{15}$\,GeV and allow $T_{\rm RH} \approx 5$\,MeV, then the correct DM abundance may be obtained by tuning the initial misalingment angle to $|\theta_i| \approx 0.1$. 
If instead $N_c = 3$, then $f_a \approx 10^{15}$\,GeV and $T_{\rm RH} \approx 10$\,MeV would require $|\theta_i| \approx 0.04$.

To discuss the properties of the glueballs, we now consider the scenario where $N_c  = 2$, $f_a \approx 10^{15}$\,GeV, $\theta_i \approx 0.1$ and $T_{\rm RH} \approx 5$\,MeV to have the correct DM abundance.
Then for $m_a \approx 0.1$\,MeV (to have a sufficiently large DM lifetime)
the dark $0^{++}$ glueball mass is $m_{0^+} \approx 2.5 \times 10^6$\,GeV. 
This implies that the dark confinement scale is $\Lambda_D \approx 4 \times 10^5$\,GeV (using the relation of the $\bar {\rm MS}$ perturbatively-computed (at 3-loop order) confinement scale to the string tension from lattice QCD in~\cite{Athenodorou:2021qvs}).    
To achieve the above reheat temperature we then need $\Lambda \approx 4 \times 10^{12}$\,GeV for $c_6 = 1$.
Referring to~\eqref{eq.lifetime}, the partial lifetime of the axion DM to two photons would then be $\sim 10^{28}$ s  for $C_{a\gamma\gamma} \sim 0.1$.  
As this example illustrates, decaying heavy axion DM, with lifetimes slightly beyond current probes, may be naturally achieved with minimal tuning due to the period of early matter domination that is necessarily generated by the dark glueballs.
\begin{figure}
    \centering
    \includegraphics[width=0.7\textwidth]{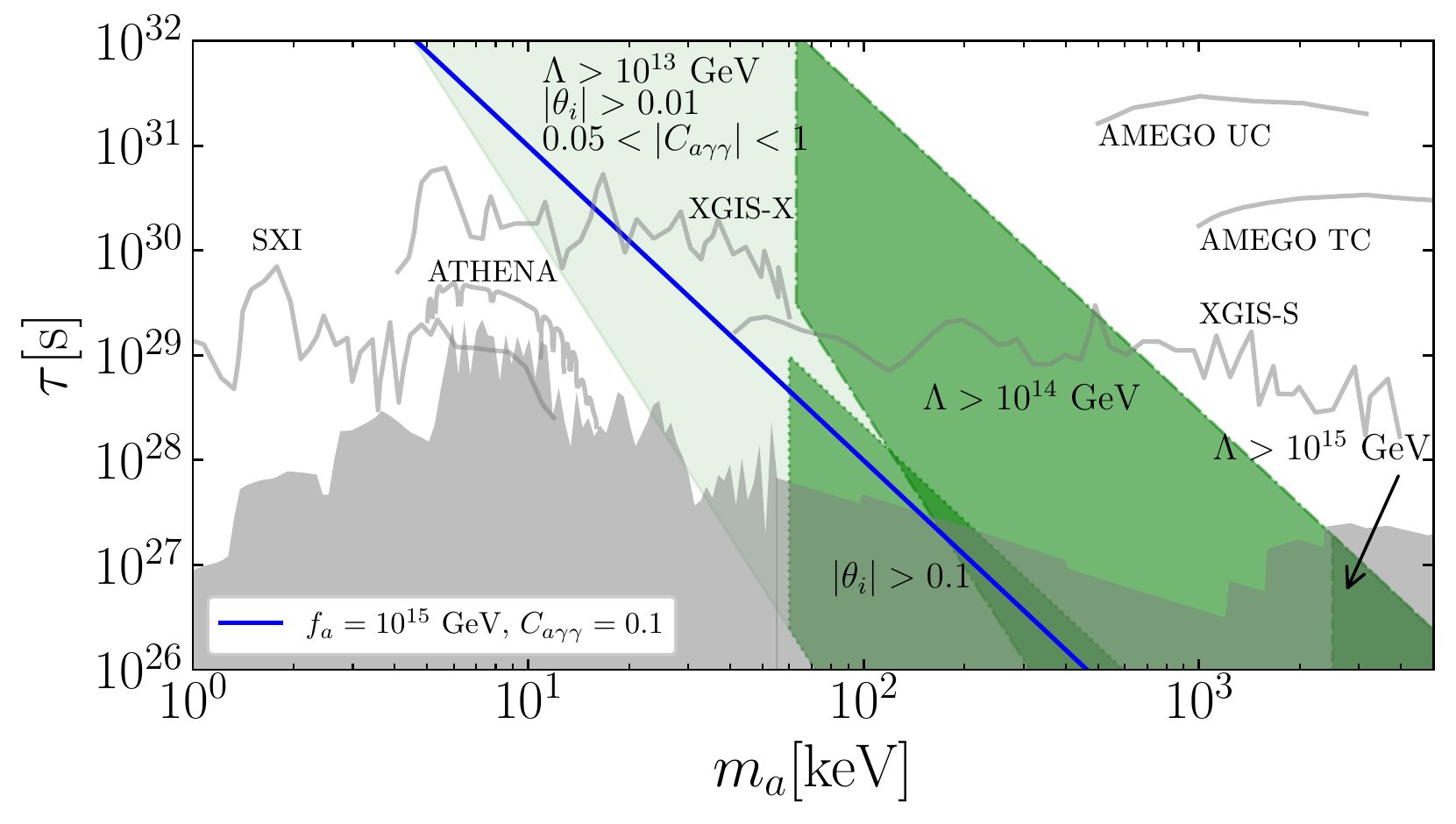}
    \caption{Parameter space for heavy axion DM. The shaded green regions illustrate where the heavy axion can constitute all of DM subject to restrictions on $\Lambda$, the scale of the dim-6 operator mediating glueball decay, $\theta_i$, and $C_{a\gamma\gamma}$, as indicated.
    The reheat temperature after glueball decay is taken to be $T_{\rm RH} > 5$~MeV to satisfy the constraint from BBN. In grey we show existing constraint from decaying DM searches, as well as projected reaches of the proposed AMEGO (with both tracked (TC) and untracked Compoton (UC) scattering), THESEUS (with instruments SXI, XGIS-X, XGIS-S), and {\it Athena} missions. We also show the axion DM lifetime as a function of its mass for a representative choice of $f_a = 10^{15}~{\rm GeV}$ and $C_{a\gamma\gamma}=0.1$.
    See text for more details.
    }
    \label{fig:lifetime_plot}
\end{figure}

In Fig.~\ref{fig:lifetime_plot} we extend this example to illustrate how the DM partial lifetime to two photons depends on the axion mass for different constraints on the initial misalignement angle, reheat temperature, axion-photon coupling, and scale $\Lambda$ that induces glueball decay.  Across the entire parameter space shaded in green we require $T_{\rm RH} > 5$ MeV, $\Lambda > 10^{13}$ GeV, $|\theta_i| > 10^{-2}$, and $0.05 < |C_{a\gamma\gamma}| < 1$, in addition to $f_a > 10^{14}$ GeV.  Note that we additionally allow $|c_6| \leq 1$, since this operator is expected to be generated at one loop.
The darker shaded green regions impose more stringent constraints, as indicated.  Note that if not otherwise stated, the constraint is the same as that described above. 
The blue line, on the other hand, shows the lifetime obtained by fixing $f_a = 10^{15}$ GeV and $|C_{a\gamma\gamma}| = 0.1$, while varying $|\theta_i|$ and $\Lambda$ to obtain the correct DM abundance at every $m_a$.

In Fig.~\ref{fig:lifetime_plot} we compare our lifetime predictions to existing constraints and projected reaches from space-based $X$-ray and gamma-ray telescopes.  The existing constraints are shaded in grey and arise from searches for $X$-ray and gamma-ray lines from (from left to right) {\it XMM-Newton}~\cite{Boyarsky:2006fg,Foster:2021ngm}, NuSTAR~\cite{Perez:2016tcq,Ng:2019gch,Roach:2019ctw}, INTEGRAL~\cite{Laha:2020ivk}, and COMPTEL~\cite{Essig:2013goa}. 
In the $X$-ray band these searches were primarily performed to search for sterile neutrino DM, which may decay into a monochromatic $X$-ray line in addition to an (unobserved) active neutrino, by looking for $X$-ray lines from DM decay in the ambient halo of the Milky Way.  
Above $\sim$ 200 keV the sensitivity to decaying DM will increase substantially in the coming years with instruments such as the AMEGO~\cite{Kierans:2020otl} and e-Astrogam~\cite{e-ASTROGAM:2017pxr} missions, which are in their planning stages.  In Fig.~\ref{fig:lifetime_plot} we show the projected reach of AMEGO to DM decaying to two gamma-ray lines from $\sim$200\,keV to $\sim$5\,MeV. AMEGO (or e-Astrogam) will improve the DM lifetime sensitivity by up to four orders of magnitude, depending on the DM mass. Our computation of the AMEGO projections is described in App.~\ref{sec:astro}.  Note that our AMEGO projections only account for statistical uncertainties to show the maximal possible science reach, though systematic uncertainties could be important and further limit the achievable lifetimes~\cite{Bartels:2017dpb}.   

In the $X$-ray band we show the projected sensitivity of the planned {\it Athena} mission, which may launch in the mid 2030's~\cite{Nandra:2013jka}, though the instrument specifications may evolve before this date.  {\it Athena} will have two instruments: the  Wide Field Imager (WFI) and the $X$-ray Integral Field Unit (X-IFU).  The X-IFU will have excellent spectral resolution ($\sim$ 5 eV versus $\sim$ 100 eV for WFI) but a smaller field of view ($\sim$ 0.014 deg$^2$ versus $\sim$ 0.7 deg$^2$ for WFI).  Both instruments will have similar effective areas (nearly a m$^2$), which are approximately an order of magnitude above those from {\it XMM-Newton}.  In fact, the WFI is comparable to the instruments onboard {\it XMM-Newton} except for the effective area.  For a search for DM decay in the ambient Milky Way halo, the signal and background fluxes are proportional to the angular size of the field of view and to the effective area, while the background flux decreases linearly with the energy resolution.  The Z-score associated with an axion signal may be estimated as $S / \sqrt{B}$ for a background-dominated search, where $S$ ($B$) is the number of signal (background) counts.  Thus, we estimate that the WFI and X-IFU instruments will have comparable sensitivity.  The sensitivity of the WFI instrument to DM decay may be roughly projected by taking the projected sensitivity to DM decay from {\it XMM-Newton} and re-scaling the lifetimes by $\sqrt{10}$ to account for the increase in effective area (assuming the same total data taking time as in the {\it XMM-Newton} analysis, which is around 30 Ms~\cite{Foster:2021ngm}).  We show this rough, projected {\it Athena} sensitivity in Fig.~\ref{fig:lifetime_plot}.   

In Fig.~\ref{fig:lifetime_plot} we also show the projected sensitivity of the THESEUS mission concept~\cite{Thorpe-Morgan:2020rwc}. THESEUS~\cite{THESEUS:2017qvx} is not an approved mission at this point but represents what may be possible in the future.  THESEUS is proposed to carry three instruments relevant for axion searches -- SXI, XGIS-X, XGIS-S -- which would collectively cover an energy range from below a keV to above an MeV.  The advantage of these instruments over, {\it e.g.}, those on {\it Athena} is the large field of view, which for THESEUS is around 1 sr across most of the energy range.  Given the comparable effective area to {\it Athena}, THESEUS would provide superior sensitivity in the mass range where the two instruments can be compared. THESEUS would also provide a transformative improvement in sensitivity near the keV scale and extend to higher masses where {\it e.g.} AMEGO would operate, though at reduced sensitivity. On the other hand, we note that the THESEUS instruments do not have improved energy resolution (with $\sim$ 200 eV resolution at a few keV).  This means that systematic uncertainties may be important for THESEUS and could ultimately limit the sensitivity in certain mass ranges. (Systematic uncertainties related to background mismodeling already limited the sensitivity of the {\it XMM-Newton} search for decaying DM in~\cite{Foster:2021ngm}, and THESEUS would have far reduced statistical uncertainties relative to those in that analysis.) Improved energy resolution is useful in part because it limits the total number of photon counts needed to achieve the target sensitivity, which means that statistical uncertainties are more important, relative to systematic uncertainties, compared to searches using telescopes that achieve the same sensitivity but with worse energy resolution.      

\section{Baryogenesis from heavy axions}\label{sec:bary}

The axion decay rate scales rapidly with the dark confinement scale $\Lambda_D$, as noted in {\it e.g.}~\eqref{eq.lifetime}; for $\Lambda_D \gtrsim 10^{10}$ GeV and GUT-scale $f_a$ the axions would decay so quickly that their cosmological abundance would be depleted before BBN.
In this section we explore the possibility that such a heavy, rapidly-decaying axion could be responsible for baryogenesis. 

For axions coupling to gauge bosons, the $B$ or the $L$ current can lead to a non-negligible baryon asymmetry in the presence of $B$ or $L$-violation through the mechanism of {\it spontaneous baryogenesis}~\cite{Cohen:1987vi}. 
Such a scenario with $L$-violation can naturally arise in the presence of the Weinberg operator, $(H \ell)^2/\Lambda_W +{\rm h.c.}$, which can explain the observed neutrino masses at the same time. 
Here $\ell $ is the left-handed lepton doublet of the SM and $\Lambda_W \sim 10^{15}~{\rm GeV}$, for which we get $m_\nu \sim 0.05$~eV (dropping flavor indices), consistent with lower bounds on the sum of neutrino masses~\cite{ParticleDataGroup:2020ssz}. 

A crucial ingredient of the spontaneous baryogenesis mechanism is coherent oscillations of (pseudo-)scalar fields, which give rise to an `effective chemical potential' for the SM fermions. 
Due to this effect, the thermal abundances of fermions and anti-fermions differ, and as a result an asymmetry between them can develop in the presence of $B$ or $L$ violation. 
In the limit of small chemical potential $\mu_i\ll T$, the asymmetry for a species $i$ is given by $\Delta n_i = n_i - \bar{n}_i \approx g_i \mu_i T^2/6$, where $g_i$ is the multiplicity of that species. 
The chemical potential induced by the scalar field, which is an axion in our applications, is determined by its coherent velocity: $\mu_i \sim \dot{a}/f_a$. Thus, the lepton asymmetry at a temperature $T$ is given by $\eta_L \propto \sum_{i=L}\Delta n_i / T^3 \sim \sum_{i=L}g_i\mu_i/T \sim \dot{a}/(T f_a)$, where the sum is over all the leptons.

The above estimate assumes that when the axion begins to oscillate, the processes mediated by the Weinberg operator are in thermal equilibrium. 
However, if axion oscillations start at temperatures $T_{\rm osc}$ lower than $T_L$, the temperature at which the Weinberg operator decouples from the bath, then the above estimate is modified to $\eta_L \sim (\Gamma_W/H(T_{\rm osc}))\times \dot{a}/(f_a T_{\rm osc})$, where $\Gamma_W \sim T^3/\Lambda_W^2$ is the rate for scattering processes through the Weinberg operator. 
In particular, in this case the produced asymmetry is suppressed by a `freeze-in'-like factor of order $\Gamma_W/H(T_{\rm osc})$. 
Given this suppression, it is clear that the produced asymmetry is maximized if the onset of axion oscillations happens at $T_L$. 
After this initial production, electroweak sphalerons convert the initial lepton asymmetry into a baryon asymmetry at the electroweak phase transition, though this processes is accompanied by a small-but-calculable efficiency factor. 

\subsection{Baryogenesis without heavy axion dark matter}

We begin by considering the possibility that there is a single heavy axion that decays before BBN and leads to baryogenesis.  
In the following subsection we generalize from this scenario to consider the possibility that the dark sector contains two confining gauge groups, leading to two massive axion states: one axion will be responsible for baryogenesis while the other will explain the DM.

To track lepton asymmetries, we study the time evolution of the chemical potential vector $\mu_i$ via the Boltzmann equation~\cite{Domcke:2020kcp}:
\begin{equation}
	\begin{aligned}
		&\frac{d}{dt}\left(\frac{\mu_i}{T}\right) = \frac{dT}{dt}\frac{1}{g_i T} \times \sum_\alpha {\cal C}_{i\alpha}\frac{\Gamma_\alpha}{H} \left(\sum_j \left(\frac{\mu_j}{T}\right){\cal C}_{j\alpha} - n_{S\alpha} \left(\frac{\dot{a}}{f_a T}\right)\right).
	\end{aligned}
	\label{eq:bary_boltz}
\end{equation}
Here $i = \tau, L_{12}, L_3, q_{12}, t, b, Q_{12}, Q_3, H$ 
runs over all the SM species, with numbers referring to SM generations.
Due to the smallness of the Yukawa couplings of the first two generations, the corresponding interactions are out of thermal equilibrium at the time of asymmetry generation.
Therefore they interact only through flavor universal gauge interactions.
Thus we can assume that the SM left-handed lepton doublets $L_1, L_2$ have the same chemical potential and denote them together as $L_{12}$. 
The same is also done for SM left-handed quark doublets $Q_1, Q_2$ and right-handed (RH) quarks $q_1, q_2$.
Along similar lines, the RH leptons of the first two generations can not interact with the bath given the absence of $SU(3)_c$ and $SU(2)_L$ interactions and the smallness of their Yukawa couplings. Thus, we need not include them.
The vector $g_i$ counts the number of degrees of freedom for different species and is given by $g_i = (1, 4, 2, 12, 3, 3, 12, 6, 4)$.\footnote{As a side-note, since the physical processes described in this section take place at a high energy scale $\sim\!10^{12}$\,GeV or even higher, it is possible that additional BSM states beyond those of the SM could be present in the thermal plasma.  
In particular, if nature realizes any form of supersymmetry below $T_{\rm osc}$ then this could lead to important quantitative and qualitative modifications to the results in this section. }

Returning to~\eqref{eq:bary_boltz}, the matrix ${\cal C}_{i\alpha}$ describes the charges of various SM species $i$ under interactions $\alpha$ and is given by,
\begin{align}
	{\cal C}_{i\alpha} = \begin{pmatrix}
	    0 & 0 & -1 & 0 & 0 & 0 & 0 \\
	    2 & 0 & 0 & 0 & 0 & 2 & 0 \\
	    1 & 0 & 1 & 0 & 0 & 0 & 2 \\
	    0 & -4 & 0 & 0 & 0 & 0 & 0 \\
	    0 & -1 & 0 & -1 & 0 & 0 & 0 \\
	    0 & -1 & 0 & 0 & -1 & 0 & 0 \\
	    6 & 4 & 0 & 0 & 0 & 0 & 0 \\
	    3 & 2 & 0 & 1 & 1 & 0 & 0 \\
	    0 & 0 & 1 & 1 & -1 & 2 & 2
	\end{pmatrix}.
	\label{eq:C_mat}
\end{align}
Here $i$ and $\alpha$ run over row and column indices, respectively. The relevant interactions $\alpha$ run over weak sphaleron, strong sphaleron, tau Yukawa, top Yukawa, bottom Yukawa, and the Weinberg operator for the first two generations and the third generation. 
As an example, consider $i=8$ which corresponds to the left-handed, third generation quark doublet $Q_3$. 
This has a non-zero charge of 3 under the weak sphaleron (three colors), 2 under the strong sphaleron (weak doublet), 0 under the tau Yukawa, 1 under both the top and bottom Yukawa, and 0 under the Weinberg operator. 
This gives a charge vector $(3, 2, 0, 1, 1, 0, 0)$, which is the eighth row of ${\cal C}_{i\alpha}$.

 The coefficients $\Gamma_\alpha$ determine the rate for the interaction $\alpha$. 
 As examples, for the dim-5 Weinberg operator ($\alpha = 6,7)$, $\Gamma_\alpha \propto T^3/\Lambda_W^2$, whereas for the marginal top Yukawa interaction ($\alpha = 4$), $\Gamma_\alpha \propto y_t^2 T$. (See App.~\ref{sec:rates} for explicit formulae for all the $\Gamma_\alpha$.) 
 
 The axion source vector $n_{S\alpha}$ depends on how the axion couples to the SM. For simplicity, we assume 
\begin{align}
	c_{aG}\left(\frac{\alpha_s}{8\pi f_a}a G\tilde{G} + \frac{\alpha_2}{8\pi f_a}a W\tilde{W} + \frac{\alpha_1}{8\pi f_a}a B\tilde{B} \right) + c_{af}\left(\sum_i \frac{\partial_\mu a}{f_a}J_i\right),
	\label{eq:axion_couplings}
\end{align}
where $J_i = \bar{f}_i\gamma^\mu f_i$ with $i$ running over all left- and right-handed SM Weyl fermions. 
Here we chose to have a single coefficient $c_{aG}$ determining all the gauge boson couplings, motivated by grand unification, and a flavor-universal coefficient $c_{af}$ for all the fermionic couplings.
We consider two benchmark choices corresponding to $c_{aG} = c_{af} = 1$ (main Article) and $c_{aG} = 1, c_{af} = 0$ (App.~\ref{app:alt}).

With this choice, we may write $n_{S\alpha} = c_{aG}( n_s + n_2) + c_{af}(\sum_i {\cal C}_{i\alpha})$. Here $n_2 = (-1, 0, 0, 0, 0, 0, 0)$ and $n_s = (0, -1, 0, 0, 0, 0, 0)$ are determined by the $aW\tilde{W}$ and  $aG\tilde{G}$ couplings, respectively. 
The term $\sum_i {\cal C}_{i\alpha}$ originates from summing over all the fermion contributions for a given interaction $\alpha$ and is determined via~\eqref{eq:C_mat} to be $(12, 0, 1, 1, -1, 4, 4)$. 
This has a vanishing entry under the strong sphaleron since QCD is a vector-like theory.
On the other hand, the three generations of left-handed quark doublets with three colors each, and lepton doublets, have a charge of $3\times 3 + 3 = 12$ under the weak sphaleron. 
Combining all these contributions, we find $n_S = c_{aG} (-1, -1, 0, 0, 0, 0, 0) +c_{af} (12, 0, 1, 1, -1, 4, 4)$.

It is useful to understand the physical effects of the various terms in~\eqref{eq:bary_boltz}. 
First we focus on the homogeneous contribution. 
The chemical potential of a species $i$ is affected by any $\alpha$ under which the species is charged. 
Furthermore, since all SM states are in thermal equilibrium, a chemical potential of species $j\neq i$ can also affect that of $i$, if $i$ and $j$ can communicate via interaction $\alpha$. 
As a toy example, if ${\cal C}_{i\alpha}$ were a diagonal $n\times n$ matrix, then~\eqref{eq:bary_boltz} would reduce to a set of decoupled homogeneous equations for each species $i$ under its exclusive interaction $\alpha$. 
The factor of $\Gamma_\alpha/H$ is a standard one denoting the efficiency of the interaction compared to the Hubble scale.

Next, we focus on the source term $n_{S\alpha}$. 
This inhomogeneous term is the one responsible for giving rise to particle-anti-particle asymmetries. 
In the absence of this term and assuming there are no initial asymmetries after inflation, we see from~\eqref{eq:bary_boltz} that $\mu_i=0$ continues to be solution at later times; {\it i.e.}, no asymmetries can develop. 
Finally, we comment on the role of the Weinberg operator, which is crucial in seeding the asymmetries in leptons in the first place. 
We consider the source term $\sum_\alpha {\cal C}_{i\alpha} (\Gamma_\alpha/H) n_{S\alpha}$ for the vector $\mu_i$. 
From this term we may derive the final $B-L$ asymmetry by first noting that
\es{}{
    \mu_{B-L} = - & (\mu_\tau + 4 \mu_{L_{12}}  + 2 \mu_{L_3})   +  (12 \mu_{q_{12}} + 3 \mu_t + 3 \mu_b + 12 \mu_{Q_{12}} + 6 \mu_{Q_3} )/3 \,.
}
Using the above and writing $\Gamma_\alpha = (\Gamma_{WS}, \Gamma_{SS}, \Gamma_\tau, \Gamma_t, \Gamma_b, \Gamma_{W_{12}}, \Gamma_{W_3})$, we find the source term for $\mu_{B-L}$ to be
$-8( \Gamma_{W_{12}} + \Gamma_{W_3})$. 
In other words, when the Weinberg operator is absent, a $B-L$ asymmetry does not get sourced, as expected.

To solve~\eqref{eq:bary_boltz} we need to know the evolution of $\dot a$ as a function of time. 
The axion dynamics, however, depend on the temperature evolution of the dark sector, since if the dark sector has an appreciable temperature then the dark axion mass may acquire non-trivial time dependence.  
We begin by considering the simpler scenario where the axion mass is temperature independent (equivalently, the dark gluons are not thermalized) before we consider the case of a temperature-dependent axion mass, as we did in Sec.~\ref{sec:DM}. 

\subsubsection{Heavy axion mass without temperature dependence}

As described above, we begin by considering the scenario where $m_a$ remains constant as the Universe evolves. 
This would be the case if the dark $SU(N)$ sector giving rise to $m_a$ was never reheated after inflation and never came into thermal equilibrium with the SM. 
In this case, the dark glueballs are not important for cosmology. 
However, along with the cold, misaligned heavy axion population with energy density $\rho_a$, there can be a relativistic axion population. 
This is because through the axion-gluon coupling, the axions can come in thermal equilibrium with the plasma if $f_a < 5\times 10^{15}{~\rm GeV}(\alpha_s/(2\pi))\left(\frac{T_{\rm RH, inf}}{10^{14}~{\rm GeV}}\right)^{1/2}$~\cite{Baumann:2016wac}. 
Here $T_{\rm RH, inf}$ is the reheat temperature after inflation.
Note that if $f_a$ is larger than this critical value there will still be a suppressed, freeze-in contribution of relativistic axions.
Such a relativistic population, with energy density $\rho_{\rm th}$, can also originate from inflaton decay.  
For example, if the inflaton has similar couplings to all SM particles and to the axion, then given the differences in degrees of freedom between the SM and the axion, we would expect $\rho_{\rm th}/\rho_{\rm SM}\sim 1/100$.  
This effectively translates into the relativistic axions having a comparable `temperature' as the SM, even if the two populations were never in thermal contact.
Therefore to be conservative, we assume $\rho_{\rm th} / \rho_{\rm SM} \simeq 1/100$, while noting that a freeze-in only production would typically give an even smaller abundance for $\rho_{\rm th}$ for large enough $f_a$ as mentioned above.
As the SM temperature $T$ falls below $m_a$, the relativistic heavy axion population starts diluting like matter and eventually decays at the same time as the cold heavy axion population. 

To track the initially generated baryon asymmetry, we therefore numerically solve~\eqref{eq:bary_boltz} along with 
\es{}{
		\ddot{a} + 3 H \dot{a} + m_a^2 a = 0 \,,
}
in addition to evolution equations for the SM plasma and $\rho_{\rm th}$ with $\rho_{\rm th} \ll \rho_{\rm SM}$ initially.
Note that the equation above is valid only for times $t$ much less than the heavy axion lifetime $\tau_a  \approx (32 \pi^3 f_a^2) / (\alpha_s(m_a)^2 m_a^3 c_{aG}^2)$. (Even in the presence of tree-level axion-matter couplings, the heavy axion with mass $m_a \gg {\rm GeV}$ would preferentially decay to gluons.)  Note also that the back-reaction of the axion-SM interactions onto the axion dynamics, predominantly arising as friction from the $SU(3)$ sphalerons, is negligible~\cite{McLerran:1990de}.   
If the SM plasma were to always dominate the energy density of the Universe then the SM energy density and the Hubble parameter would evolve as
\es{eq:SM_density_evolution}{
	\dot{\rho}_{\rm SM} + 4 H \rho_{\rm SM} \approx 0,\\
	\rho_{\rm SM} \approx 3 H^2 \mpl^2 \,.
}
However, the axions can come to dominate the energy density at later times, and their eventual decays would in this case dilute the initially-generated baryon asymmetry. 
To compute this dilution we do not solve~\eqref{eq:SM_density_evolution} but rather the more general set of equations
\es{}{
&\dot{\rho}_a + 3H \rho_a + \frac{\rho_a}{\tau_a} = 0 \,, \\
&\dot{\rho}_{\rm th} + 4H \Theta(T - m_a) \rho_{\rm th} + 3H \Theta(m_a -T) \rho_{\rm th} + \frac{\rho_{\rm th}}{\tau_a} = 0,\\
&\dot{\rho}_{\rm SM} + 4 H \rho_{\rm SM} - \frac{\rho_a}{\tau_a} - \frac{\rho_{\rm th}}{\tau_a} = 0,\\
&3 H^2 \mpl^2 = \rho_{\rm SM} + \rho_a +\rho_{\rm th},\\
&\dot{\Delta n_B} + 3 H \Delta n_B = 0.
}
In the second line we use the approximation that the relativistic axion population $\rho_{\rm th}$ instantaneously transitions from $1/R^4$ dilution to $1/R^3$ dilution at $T=m_a$ and denote this with the unit step function $\Theta(x)$. 
Before the entropy diluton, the `initial' baryon asymmetry $\Delta n_B$ is given by, 
\begin{align}
	Y_B = \frac{\Delta n_B}{s} = -c_{\rm sph}\frac{\Delta n_{B-L}}{s} =  -c_{\rm sph}\frac{\mu_{B-L} T^2}{6s}.
\end{align}
Here we normalize the baryon asymmetry with respect to the entropy density $s$, and we use a sphaleron conversion factor $c_{\rm sph} \approx 0.35$ to convert the $B-L$ asymmetry into a $B$ asymmetry at the electroweak phase transition~\cite{Harvey:1990qw}. 

\begin{figure*}
	\begin{center}
		\includegraphics[width=0.49\textwidth]{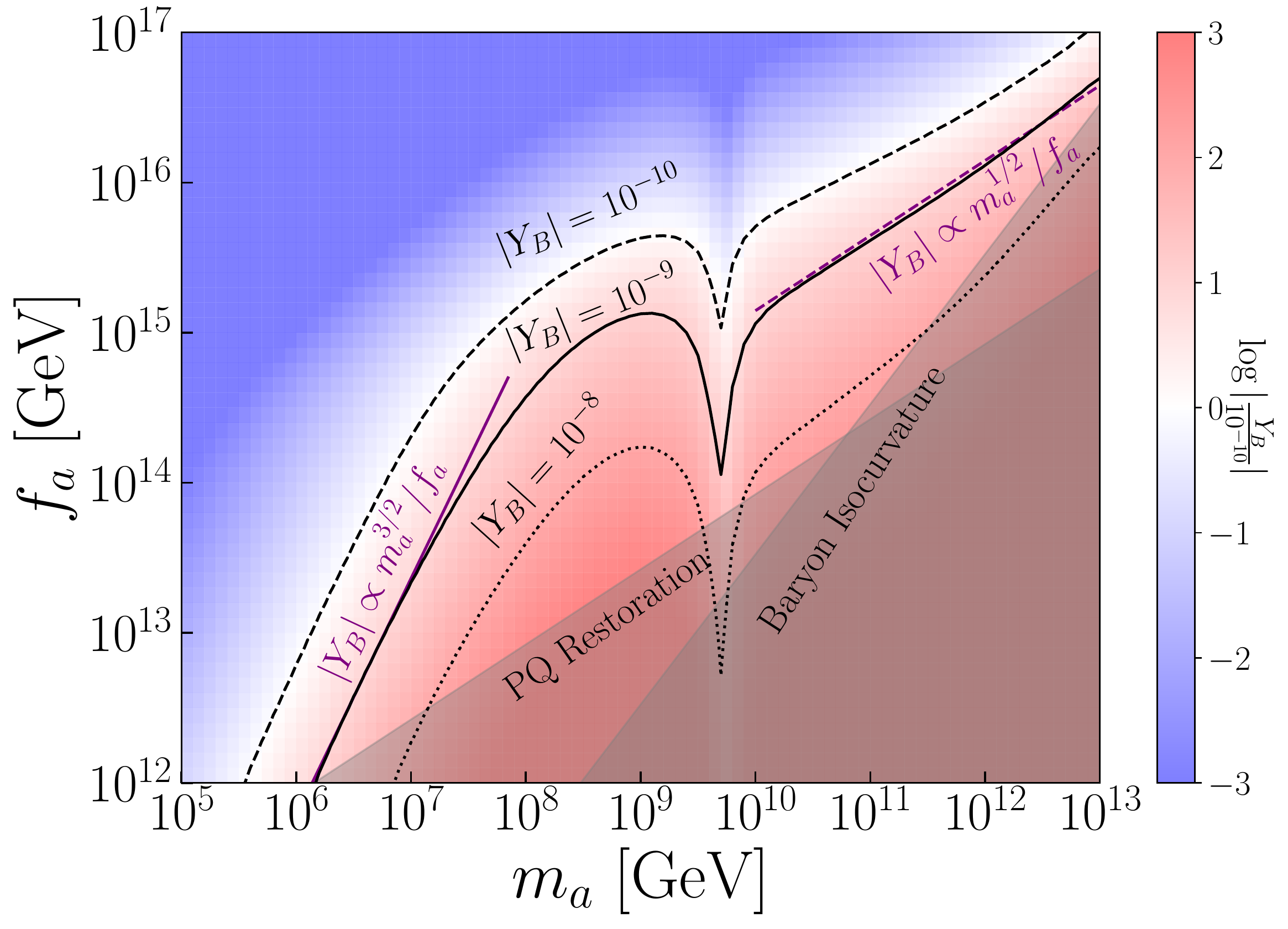}
		\includegraphics[width=0.49\textwidth]{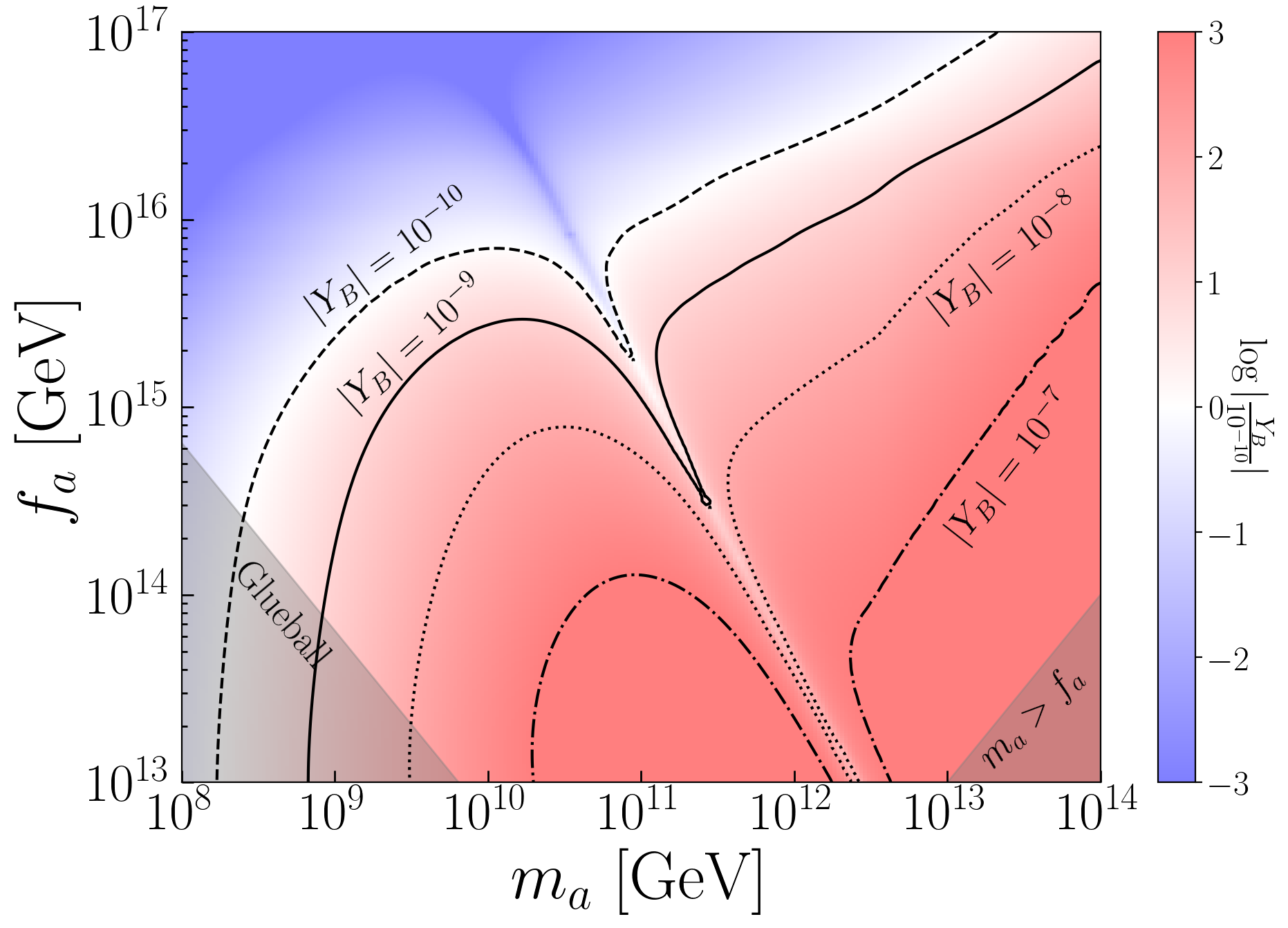}
	\end{center}
	\caption{The primordial baryon asymmetry $Y_B$ generated from spontaneous baryogenesis due to heavy axion oscillations for axions coupled to a confining dark sector. (Left) We assume that the heavy axion mass $m_a$ is temperature-independent. (Right) We assume that mass is temperature dependent for an $SU(3)$ dark sector that has the same temperature as the SM.  In this figure we assume $c_{aG} = c_{af} = 1$, though $Y_B$ scales linearly with the couplings.  Red regions overproduce the baryon asymmetry while blue regions underproduce it. The overproduction regions are still important because they can allow for extra entropy dilution at late times, which is the case when we include an additional, lighter axion that explains the DM. Note that the dips in the contours arise from an accidental cancellation in the lepton asymmetries generated between the axion-gauge couplings and the axion-matter couplings in~\eqref{eq:axion_couplings}; this cancellation would generically not occur for non-universal couplings, as illustrated in App.~\ref{app:alt}.  Note that in the right panel we chose $c_6 = 1$ and $\Lambda = 10^{13}$\,GeV -- parameters which control the glueball decay -- for definiteness. 
	}
	\label{fig:heavy_ax_T_ind}
\end{figure*}

Since the heavy axion would have its own quantum fluctuations during inflation, it would source baryon isocurvature fluctuations, which are constrained by the Planck mission~\cite{Planck:2018vyg,Planck:2018jri}. 
Those constraints translate to
\begin{align}
	\frac{H_{\rm inf}}{|\theta_i| f_a} \lesssim 3.2\times 10^{-4} \,,
	\label{eq:bary_iso}
\end{align}
where $\theta_i$ is the initial misalignment angle of the heavy axion and $H_{\rm inf}$ is the Hubble scale during inflation.
We also need to have $m_a < H_{\rm inf}$ so that the axion misalignment angle is not driven to zero during inflation. 
Combining the two equations above we find the constraint
\begin{align}
	m_a \lesssim 3.2\times 10^{-4} |\theta_i| f_a \,.
\end{align} 
The resulting constraint is labelled as `Baryon Isocurvature' in Fig.~\ref{fig:heavy_ax_T_ind}.

We also require that the energy density in the misaligned heavy axion population is smaller than the SM bath at the inflation reheat temperature, $T_{\rm RH,inf}$, otherwise axions would dominate the energy density during inflation. 
This requirement translates to, 
\begin{align}
	\frac{1}{2}m_a^2 (\theta_i f_a)^2 \ll \frac{\pi^2}{30}g_* T_{\rm RH, inf}^4.
	\label{eq:axion_subdom}
\end{align}
The requirement that the effective description of the dimension-5 Weinberg operator is valid implies $T_{\rm RH, inf}<\Lambda_W$, where as mentioned $\Lambda_W \sim 10^{15}$~GeV to achieve $m_\nu \sim 0.05$~eV. Thus, we require
\begin{align}
	T_{\rm RH, inf} < 10^{15}\,\GeV  \,,
\end{align}
though if $T_{\rm RH,inf}$ is near or above this scale it may serve to increase the baryon asymmetry by the standard thermal leptogenesis mechanism of decaying right-handed neutrinos~\cite{Davidson:2008bu}.
A more constraining requirement for lighter masses comes from demanding that the PQ symmetry that produces the heavy axion is not restored after inflation; {\it i.e.}, $T_{\rm RH, inf} < f_a$.  
If the PQ symmetry is restored then $\langle \dot a \rangle = 0$, averaged over large (super-horizon) scales, in which case no coherent baryon asymmetry is generated.  
We require
\es{}{
    T_{\rm RH, inf} > 6 \times 10^{12} \, \, {\rm GeV} \,,
}
as otherwise the Weinberg operator would never be in thermal equilibrium~\cite{Domcke:2020kcp}, which could significantly the suppress the generated baryon asymmetry. 
To map this constraint onto the $m_a$-$f_a$ parameter space, we assume efficient reheating after inflation and set
\begin{align}
	H_{\rm inf} \mpl \approx \frac{\pi}{\sqrt{90}}\sqrt{g_*}T_{\rm RH, inf}^2.
\end{align}
Then combining the restrictions $m_a < H_{\rm inf}$ and $T_{\rm RH, inf} < f_a$, we arrive at
\begin{align}
    m_a < \frac{\pi}{\sqrt{90}} \sqrt{g_*} \frac{f_a^2}{\mpl}.
\end{align}
This constraint is labelled as `PQ Restoration' in Fig.~\ref{fig:heavy_ax_T_ind}. 
The other constraint that is important for validity of the axion effective field theory (EFT) is $T < f_a$, where $T$ is the temperature where the baryon asymmetry is dominantly generated. However, with the stronger constraint of $T_{\rm RH, inf} < f_a$, this restriction is already obeyed. This also ensures that the backreaction of the produced charges on the axion dynamics is small.

Contours for various values of the present-day baryon abundance, subject to the above constraints, are illustrated in  Fig.~\ref{fig:heavy_ax_T_ind} (left) for $c_{aG} = c_{af} = 1$ along with $\theta_i = 1$. 
Note that $\theta_i$ may be larger than unity, which would increase the baryon asymmetry, though then anharmonic effects may become important, as we discuss further below.  
For large regions (colored white) of parameter space we produce the correct, observed baryon asymmetry.  
Blue regions underproduce the baryon asymmetry, while red regions correspond to overproduction.  
The red regions will play an important role when adding in the second, DM axion, as this sector will add additional entropy dilution that can dilute the red regions to the observed baryon asymmetry. 
Note that for $m_a \sim 10^{10}$\,GeV the baryon asymmetry contours acquire a sharp dip.  
This dip arises partially because of a cancellation 
between axion-gauge-coupling produced asymmetry and the axion-matter produced asymmetry;
the dip, while still present, is less pronounced in the figures in App.~\ref{app:alt} that have $c_{aG} = 1, c_{af} = 0$. 

We can also intuitively understand the shapes of constant $|Y_B|$ contours for smaller values of $m_a$, to the left of the dip. In this regime, the asymmetry production typically happens in the `freeze-in' regime with the initial lepton asymmetry $\eta_L \sim (\Gamma_W/H(T_{\rm osc}))\times \dot{a}/(f_a T_{\rm osc}) \propto m_a$, with no dependence on $f_a$. If not for the heavy axion domination, the baryogenesis contours would then have been horizontal. However, in this parameter space, initially thermal heavy axions do come to dominate the energy density of the Universe at $T_{\rm dom} \sim m_a$, while they decay at $T_{\rm RH} \propto m_a^{3/2}/f_a$. Therefore, the final abundance scales as $m_a\times m_a^{1/2}/f_a$. On the other hand, for larger values of $m_a$, to the right of the dip, the axion is already oscillating when the Weinberg operator decouples. As a result the initial asymmetry is mildly dependent on $m_a$. However, the entropy dilution is the same as before and hence the final abundance scales as  $m_a^{1/2}/f_a$. We show these two parametric expectations, for small and large $m_a$, by solid and dashed purple lines, respectively.

\subsubsection{Heavy axion mass with temperature dependence}

Now we consider the scenario where all the relevant degrees of freedom are reheated after inflation. 
This implies along with the SM bath, there is also a thermal population of heavy axions, dark gluons, and finally, a cold, misaligned heavy axion population as before. 
We  focus on the part of parameter space where dark gluons decay into the SM soon after dark confinement. 
This ensures that the generated baryon asymmetry is not diluted due to heavy glueball domination.
To ensure the glueballs promptly decay we require $\Gamma_{0^{++} \to {\rm SM}} > H(T_c)$, where $T_c$ is the confinement temperature and $\Gamma_{0^{++} \to {\rm SM}}$ the glueball decay rate~\eqref{eq:glueball_decay}. 
This implies that
\begin{equation}
	\begin{aligned}
		& 6\times 10^{-5} c_6^2 \frac{x^5 (m_a f_a)^{5/2}}{\Lambda^4} > \frac{\pi\sqrt{g_*}}{\sqrt{90}\mpl}m_a f_a\left(1.6 - \frac{0.8}{N_c^2}\right)^2 \\
		& \Rightarrow m_a f_a > {3\times 10^{25}{~\rm GeV}^2 \over c_6^{4/3}} {\left(\Lambda\over10^{14}{~\rm GeV}\right)}^{8/3}~{\rm for}~N_c = 3.
	\end{aligned}
	\label{eq:h_glueball_decay}
\end{equation}
In the last relation above we specify to the case of a dark $SU(3)$ gauge group that is responsible for the heavy axion mass. 
We also take $x \equiv m_{0^+} / \sqrt{m_a f_a}\approx 8$, as before.

The equations governing the generation and evolution of the baryon asymmetry are the same as in the previous subsection, except that now $m_a \rightarrow m_a(T)$ for $T>T_c$, with the temperature-dependent mass as given in Sec.~\ref{sec:DM} (we assume $SU(3)$ for definiteness).  
For simplicity, we also assume that the SM and dark sectors have the same temperature, though in principle the dark sector could be either colder or hotter than the SM if the two were not in equilibrium or reheated differently after inflation. The result is shown in Fig.~\ref{fig:heavy_ax_T_ind} (right).

The baryon isocurvature constraint in~\eqref{eq:bary_iso} applies as before. 
However, since during inflation the dark sector is deconfined, we have $m_a(T)\ll H_{\rm inf}$, and thus the isocurvature constraint becomes independent of the axion mass. 
The vanishing axion mass also ensures that the energy density in axions is always subdominant during inflation. 
The restriction due to $T_{\rm RH, inf} < f_a$ continues to apply and is also independent of $m_a$.
We also show the region labeled $m_a > f_a$ where the axion cannot be treated as a light Goldstone boson. 
Finally, the constraint from~\eqref{eq:h_glueball_decay} shows that in the shaded region labeled `Glueball', the heavy glueballs do not decay promptly. 
Consequently, the entropy dilution coming from their decay needs to be taken into account in this parameter space, which we have not done for simplicity. 
Therefore in that region, our computation of $Y_B$ does not apply. 
We also have chosen $\Lambda = 10^{13}$\,GeV with $c_{aG} = c_{af} = 1$ as an illustration, along with $\theta_i = 1$ (though see App.~\ref{app:alt}).  
Increasing $\Lambda$ makes the glueballs longer lived, which may further dilute the baryon asymmetry if the glueballs come to dominate the energy density.  Lastly, note that in all of the parameter space illustrated in Fig.~\ref{fig:heavy_ax_T_ind} the $\mu/T$ values are smaller, by at least a few orders of magnitude, compared to those recently constrained in~\cite{Domcke:2022uue} by helical magnetic field generation.

\subsection{Heavy axion baryogenesis and light axion dark matter}

We now focus on a scenario where there are two axions in the spectrum: a heavy axion $a_h$ and a light axion $a_l$. 
They get their masses from dark $SU(N_h)$ and $SU(N_l)$ groups, respectively. 
We consider $N_l < N_h$, such that $a_h$ is heavier than $a_l$ for similar decay constants.
We assume both the sectors are reheated after inflation.
Therefore, at reheating we have the following populations: 
(a)~cold, misaligned population of both $a_h$ and $a_l$ ($\rho_a^h, \rho_a^l)$; 
(b)~a relativistic population of both $a_h$ and $a_l$ ($\rho_{\rm th}^h, \rho_{\rm th}^l$); and 
(c)~deconfined $SU(N_h)$ and $SU(N_l)$ gluons ($\rho_G^h, \rho_G^l)$.  
The goal of this subsection is to explore the parameter space for which the two axions can explain both the DM relic density and the primordial baryon asymmetry. 
This is non-trivial because the same early matter dominated era that is required to avoid DM overclosure dilutes the already generated baryon asymmetry.

As in the previous section, we consider the parameter space where the $SU(N_h)$ glueballs decay soon after their confinement since they would otherwise give rise to a very early matter domination with subsequent entropy dump that would dilute the initial baryon abundance.
This requirement is the same as in~\eqref{eq:h_glueball_decay}. 

The early cosmological history in this scenario proceeds as follows. 
After inflation, the Universe becomes radiation dominated with the thermal bath consisting of relativistic axion populations, the deconfined dark plasmas, and the SM. 
We assume all of these to have the same temperature for simplicity.
When $H \sim m_h$, the field $a_h$ starts to oscillate and this generates a lepton asymmetry in the presence of the Weinberg operator.
At $T < m_h$, $\rho^h_{\rm th}$ starts diluting like matter. 
Together with $\rho_a^h$, these cold populations can give rise to matter domination if they are sufficiently long lived. 
Then at the heavy axion lifetime $\tau_a(m_h, f_h)$, both $\rho_a^h$ and $\rho^h_{\rm th}$ decay. 
We assume that heavy axion decay contributes equally to the SM and $SU(N_l)$ gluons in terms of energy density. 
At times immediately after the heavy axion decay the Universe remains radiation dominated with $\rho_{\rm SM}, \rho^l_{\rm th}, \rho_{G}^l$. 
At $T<T_{c}^l$, $SU(N_l)$ confinement takes place, and subsequently $\rho_G^l$ gives rise to glueballs that soon start dominating the energy density. 
This gives rise to a matter-dominated epoch. 
These glueballs eventually decay before BBN and reheats the Universe.
Following this point the evolution is same as in standard cosmology.
When $T<m_l$, the $\rho_{\rm th}^l$ also start diluting like matter and these warm axions can potentially form a sub-component of DM.
 
With this cosmology in mind, we now ask for which parameter space we get the correct DM and baryon abundances. 
Consider a heavy axion with $m_h = 10^{11}$\,GeV and $f_h = 5 \times 10^{13}$\,GeV, along with $\theta_i^h = 1$, $c_{aG} = c_{af} = 1$ and $T_{\rm RH, inf} = 10^{13}$~GeV. 
This implies that gluons of the heavy sector confine around $T\approx 4\times 10^{12}$\,GeV to form heavy glueballs. 
However for $\Lambda \sim 10^{13}$~GeV, these glueballs decay promptly after their production, as implied by~\eqref{eq:h_glueball_decay}. 
Since the dark gluon sector is assumed to have the same temperature as the SM, their energy density is $2(N_c^2-1)/g_* \sim 1/10$ of the SM. 
Consequently, the heavy glueball formation and their prompt decay does not affect the thermal bath significantly. 
We now take the light axion parameters to be $m_l = 7$\,keV and $f_l = 7\times 10^{13}$\,GeV. 
This implies that the second dark confinement transition happens around $T\approx 30$\,TeV, following which lighter dark glueballs with mass $2 \times 10^5$\,GeV form and soon come to dominate the energy density.
Through the dimension-6 Higgs portal coupling, these glueballs eventually decay. 
Taking $c_6 =1 $
and $\Lambda = 5\times 10^{9}$\,GeV, we compute the corresponding reheat temperature to be $T_{\rm RH}\approx 3$\,GeV using~\eqref{eq:TRH_glue}.
The entropy dilution caused by the glueball decay dilutes the initial value of the baryon asymmetry, and with the above choices of parameters we find $|Y_B| \approx 10^{-10}$, consistent with current observations. 
Using~\eqref{eq:Tosc_over_Tc} for $N_c=2$, we obtain $T_{\rm osc} \sim 3.5 \times 10^{5}$\,GeV. 
Given that the onset of matter domination happens around $T_{\rm EMD}\sim 30$\,TeV, the observed DM density can be explained for $|\theta_i| \sim 10^{-2}$ using~\eqref{eq:mod_relic_ab}. Lastly, for $C_{a\gamma\gamma} \sim 0.05$, the DM lifetime is determined to be $6\times 10^{29}$\,sec using~\eqref{eq.lifetime}, consistent with current searches for decaying DM, but can be probed with {\it Athena} or THESEUS.

\begin{figure*}
    \begin{center}
    \includegraphics[width = 0.48 \textwidth]{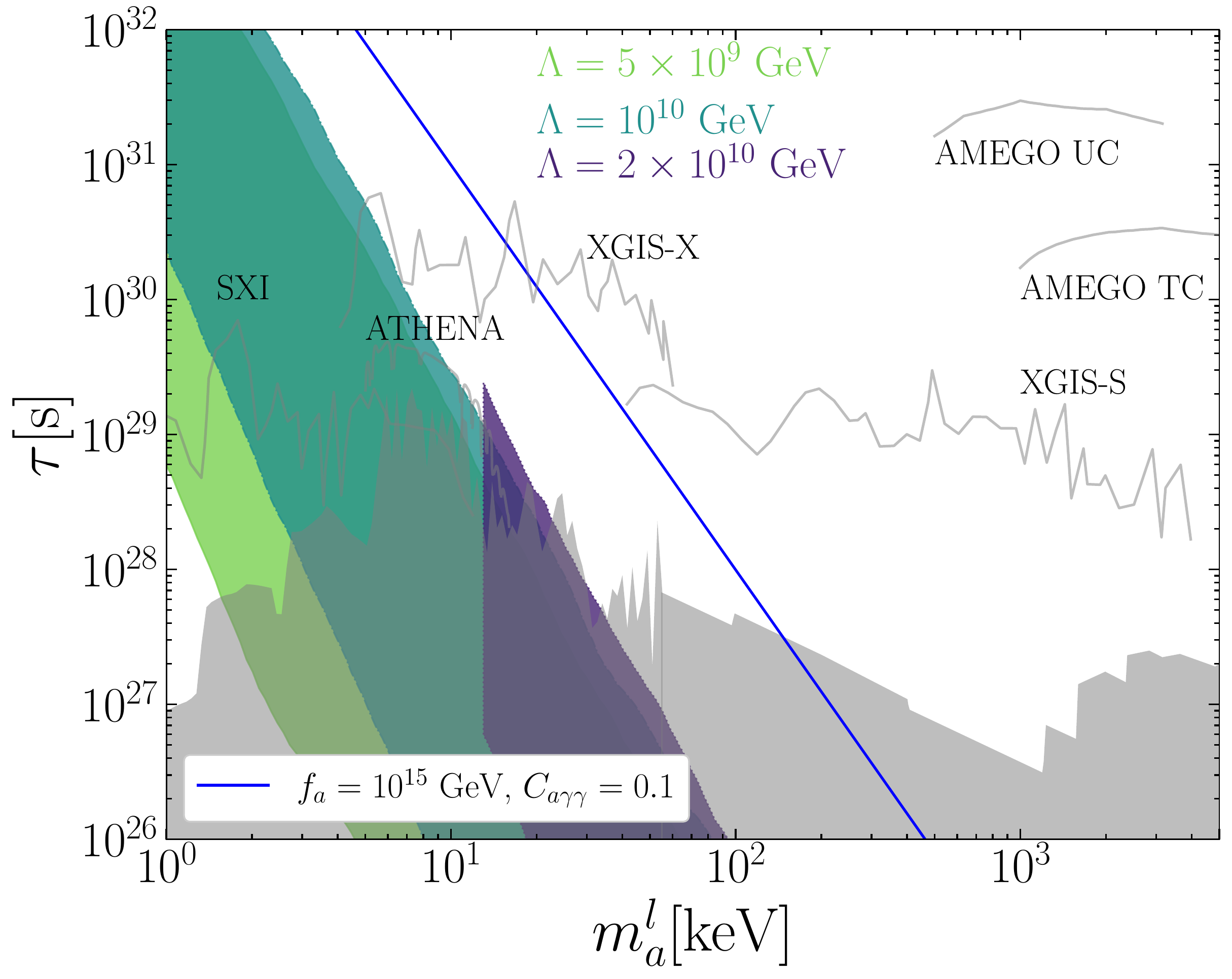}
    \includegraphics[width = 0.48 \textwidth]{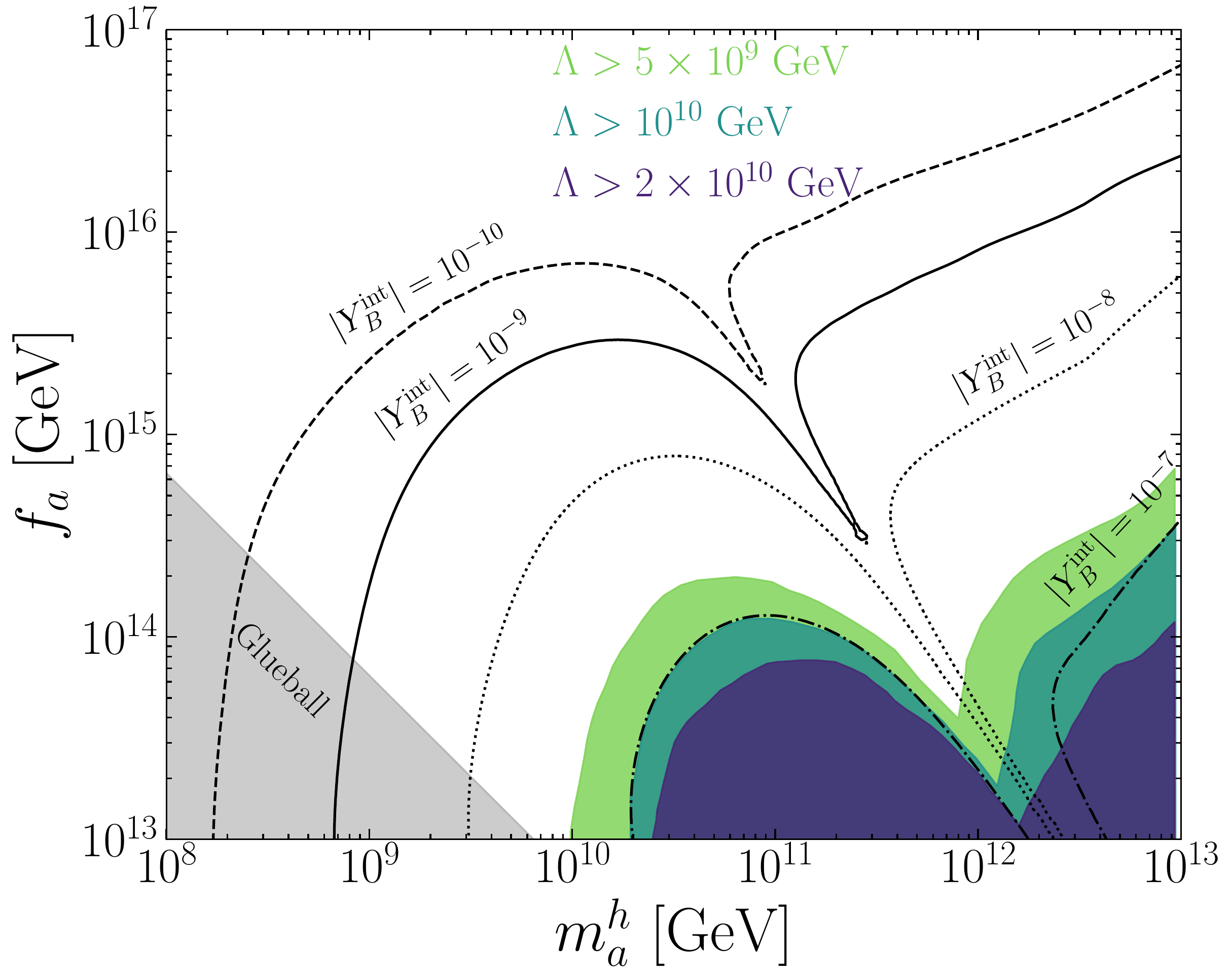}
    \end{center}
    \caption{(Left) As in Fig.~\ref{fig:lifetime_plot} but for the case where there is both a keV-MeV axion, illustrated in the left panel, that explains the primordial DM and a much heavier axion, illustrated in the right panel, that explains the primordial baryon asymmetry through spontaneous baryogenesis.  We illustrate lifetimes and masses where these two abundances are correctly reproduced for fixed values of the intermediate scale $\Lambda$. Note that unlike in Fig.~\ref{fig:lifetime_plot} the $\Lambda$ are taken to be well below the GUT scale, which is necessary to avoid over-diluting the baryon abundance. As in Fig.~\ref{fig:lifetime_plot} we allow $|\theta_i^l| >0.01$ and $0.05 < |C_{a\gamma\gamma}| < 1$ while requiring $T_{\rm RH} > 5$ MeV.  (Right) The heavy axion parameter space in the two-axion model, illustrated by the heavy axion mass and its decay constant.  Shaded regions show where the two primordial abundances are correctly produced for $\Lambda$ greater than the indicated values.  The $Y_B^{\rm int}$ contours show how much `intermediate' baryon asymmetry is produced before the entropy dilution through glueball domination and reheating of the DM dark sector ({\it i.e.}, the dark sector with the lower confining scale). These contours are the same as in Fig.~\ref{fig:heavy_ax_T_ind}.  In this case we must overproduce the baryon asymmetry at intermediate scales, since the period of early matter domination, which is necessary to achieve the correct DM abundance, dilutes the baryon asymmetry.}
    \label{fig:tau_bary}
\end{figure*}

In Fig.~\ref{fig:tau_bary} we extend the above argument for a broader parameter space,
highlighting the appropriate regions of parameter space for which the correct baryon and DM abundances are achieved.  We fix $\theta_i^h = 1$, constrain $|\theta_i^l| > 10^{-2}$, and consider $\Lambda > 5 \times 10^9$ GeV along with $\Lambda > 2 \times 10^{10}$ GeV, fixing $c_6 = 1$.  
We allow the two axions to have different decay constants so long as they are above $10^{13}$ GeV. As in Fig.~\ref{fig:lifetime_plot}, we vary $0.05 < |C_{a\gamma\gamma}| < 1$.  In the left (right) panel we illustrate the light (heavy) axion parameter space where the correct DM and baryon abundances are simultaneously obtained. In the left panel we show the lifetime to photons instead of $f_a$, since this is directly observable, as a function of the light axion mass, while in the right we show $f_a$ as a function of the heavy axion mass.  Note that the preferred mass range for the DM axion is lower than in Fig.~\ref{fig:lifetime_plot}.
We also note that the viable parameter space in Fig.~\ref{fig:tau_bary} left is not strictly nested as $\Lambda$ is increased, contrary to Fig.~\ref{fig:tau_bary} right. We label the left panel for fixed, illustrative values of $\Lambda$.

In Fig.~\ref{fig:tau_bary} we fix $\theta_i^h = 1$, though in principle $\theta_i^h$ could be larger, which may enhance the baryon abundance and thus open up more of the DM parameter space where the simultaneous DM and baryon abundances may be reproduced. In particular, it is possible that $\theta_i^h$ could be near $\pi$, in which case anharmonicities in the heavy axion equation of motion become important.  In particular, for $\theta_i^h = \pi - \delta_i$, with $\delta_i \ll 1$ a small, positive number, it is known that the heavy axion field value becomes logarithmically enhanced in $\delta_i$ at late times (see, {\it e.g.},~\cite{Visinelli:2009zm}).  However, since the baryon abundance is at most logarithmically  enhanced as $\theta_i^h$ is tuned towards $\pi$,  anthropic selection of $\theta_i^h$ near $\pi$ to enhance the baryon abundance may not be efficient, though this deserves further consideration.

We note that the scale $\Lambda$ controlling the glueball decay rate needs to be much smaller than $M_{\rm GUT}$ for successful baryogenesis to occur. 
In the next section, we describe an example UV completion that achieves $\Lambda \ll M_{\rm GUT}$.  

\section{Orbifold construction}
\label{sec:model}

So far in this Article we have motivated the scale $\Lambda_D$ by assuming a unified gauge group at some scale $M_{\rm GUT}\gtrsim 10^{16}$\,GeV that breaks to $G_{\rm SM}\times G_{\rm dark}$ below that scale. 
We now give an example, extra dimensional construction that achieves such a breaking pattern. 
To be concrete, we focus on orbifold GUTs and consider unification of $G_{\rm SM}$ with $SU(3)_D$. 
Construction with more general $SU(N)_D$ or $SU(N)_D \times SU(M)_D$ can be carried out in a similar way.

Orbifold GUTs are extra dimensional constructions that explain grand unification in a simple and elegant way. 
The basic idea is that in the presence of compact extra dimensions, one needs to specify boundary conditions to completely describe the theory. 
It is these boundary conditions that can break the unified gauge group and also project out the unwanted zero-modes of various fields, avoiding issues such as proton decay and the doublet-triplet splitting problem.
We now briefly review some necessary aspects of an orbifold construction while referring the reader to~\cite{Kawamura:2000ev,Hall:2001pg,Hebecker:2001wq} for more details.

We consider the spacetime to be $M_4\times S_1/(Z_2\times Z_2')$ where $M_4$ denotes the 4D Minkowski spacetime, with coordinates denoted by $x$. 
The extra dimensional circle $S_1$ with radius $R$ is reduced to an interval due to the quotienting by $(Z_2\times Z_2')$. 
Here the first $Z_2$ implements an identification $y\rightarrow -y$ where $y$ is the coordinate along the extra dimension. 
The second identification, $Z_2'$ acts as $y'\rightarrow-y'$ where $y'=y-\pi R/2$, or equivalently, $y\rightarrow \pi R-y$. 
The action of both of these parity transformations restricts the original $y$ coordinate ranging from $0\leq y < 2\pi R$, to $0\leq y \leq \pi R/2$, with the rest of the circular space identified to this segment. 
In particular, the end points $y=0$ and $y=\pi R/2$ act as orbifold fixed points where other fields, such as those in the SM, can be located. 
We denote the parity transformations associated with $Z_2$ and $Z_2'$ as $\cal P$ and $\cal P'$, respectively. 
In particular, focusing on an $SU(N)$ gauge field in the bulk, $\cal P$ has an action,
\begin{equation}
\begin{aligned}
{\cal P}: A_\mu(x,y) &\rightarrow A_\mu(x,-y)  = P A_\mu(x,y) P^{-1},\\
{\cal P}: A_5(x,y) &\rightarrow A_5(x,-y)  = -P A_5(x,y) P^{-1},\\
\end{aligned}
\end{equation}
where $P$ is an $N\times N$ matrix with eigenvalues $\pm 1$. 
The action of $\cal P'$ is defined analogously via a matrix $P'$. 
We note that under the action of a given parity operation, $A_\mu$ and $A_5$ transforms oppositely, as needed for invariance of the Lagrangian. 
For a field $\Phi$ in the fundamental of $SU(N)$,
the actions of $\cal P, \cal P'$ are given by,
\begin{equation}
\begin{aligned}
{\cal P}: \Phi(x,y) &\rightarrow \Phi(x,-y)  = P \Phi(x,y),\\
{\cal P'}: \Phi(x,y') &\rightarrow \Phi(x,-y')  = P' \Phi(x,y').\\
\end{aligned}
\end{equation}
To determine the action of $\cal P, \cal P'$ it is useful to recall the mode expansion of a bulk field $\phi(x,y)$ that has specific parity properties (see, {\it e.g.}, \cite{Hall:2001pg}),
\begin{equation}
\label{eq:extra_dim_mode}
\begin{aligned}
\phi_{++}(x,y) & = \sum_{m=0}^{\infty}\frac{1}{\sqrt{2^{\delta_{m,0}}\pi R}}\phi_{++}^{(2m)}(x)\cos(2my/R),\\
\phi_{+-}(x,y) & = \sum_{m=0}^{\infty}\frac{1}{\sqrt{\pi R}}\phi_{+-}^{(2m+1)}(x)\cos((2m+1)y/R),\\
\phi_{-+}(x,y) & = \sum_{m=0}^{\infty}\frac{1}{\sqrt{\pi R}}\phi_{-+}^{(2m+1)}(x)\sin((2m+1)y/R),\\
\phi_{--}(x,y) & = \sum_{m=0}^{\infty}\frac{1}{\sqrt{\pi R}}\phi_{--}^{(2m+2)}(x)\sin((2m+2)y/R).
\end{aligned}
\end{equation}
Here the notation, $\phi_{++}$ for example, implies that the field is even under both $\cal P, P'$. The fields $\phi_{++}^{(2m)}, \phi_{+-}^{(2m+1)}, \phi_{-+}^{(2m+1)}, \phi_{--}^{(2m+2)}$ have masses $2m/R, (2m+1)/R, (2m+1)/R, (2m+2)/R$, implying only $\phi_{++}$ has a zero-mode (setting $m=0$) and is present in the low-energy EFT below the scale $1/R$. 

To recall how gauge coupling unification works in this scenario, consider the action for a bulk gauge theory in flat spacetime, 
\begin{align}
    S \supset \int_0^{\pi R} dy \int d^4 x 
    \left(\frac{1}{g_5^2}F_{AB}F^{AB} + \delta(y) \sum_i \epsilon_i F_{i,\mu\nu} F_i^{\mu\nu}\right) \,.
\end{align}
The 5D gauge coupling is $g_5$, and we assume that the bulk gauge invariance is broken at the $y=0$ boundary.  
Consequently, we can write non-GUT symmetric contributions to individual gauge groups parameterized by $\epsilon_i$. 
The indices $A,B$ run over all the dimensions whereas $\mu,\nu$ run over only 4D. The zero modes of the gauge bosons have a flat profile in the extra dimension, as can be seen from~\eqref{eq:extra_dim_mode}. 
Integrating over the extra dimension we then find at the unification scale,
\begin{align}\label{eq:couling_gut}
    \frac{1}{\alpha_i(\mu \simeq 1/ R)} = \frac{4\pi^2 R}{g_5^2} + 4\pi \epsilon_i \,,
\end{align}
where we match the value of $\alpha_i$ at the renormalization scale $\mu = 1/R$ to the 5D coupling.
This implies as long as the size of the extra dimension is large, {\it i.e.}, $\pi R/g_5^2 \gg \epsilon_i$, all the gauge couplings $\alpha_i$ are unified at the scale $1/R$, while below that scale, each $\alpha_i$ has their own evolution.\footnote{Above the compactification scale $1/R$, there can also be some small differential running of the gauge couplings since Kaluza-Klein modes of bulk fields may not a fill an entire gauge multiplet. In this case \eqref{eq:couling_gut} would approximately hold with a precise unification taking place somewhat above $1/R$. See, {\it e.g.}~\cite{Hall:2001pg, Nomura:2001mf}.}

\subsection{Orbifold construction of $SU(6)\rightarrow SU(3)_D \times SU(3)_c$}

First we consider a warm up example in which only QCD is unified with $SU(3)_D$ but $SU(2)_L\times U(1)_Y$ does not unify.
We imagine an extra dimensional scenario with $S_1/(Z_2\times Z_2')$ geometry, as described above. The bulk gauge group is $SU(6) \times SU(2)_L \times U(1)_Y$.

For the boundary at $y=0$, we choose $P={\rm diag}(1, 1, 1, -1, -1, -1)$, whereas for $y=\pi R/2$, we choose $P'={\rm diag}(1, 1, 1, 1, 1, 1)$. With this choice, $SU(6)\rightarrow SU(3)_D \times SU(3)_c \times U(1)$ on the $y=0$ boundary, whereas the bulk gauge invariance remains intact on the $y=\pi R/2$ boundary.
This shows that the low energy theory has a $SU(3)_c\times SU(3)_D\times U(1)$ symmetry. 
We index the unbroken generators by $a$ and the broken ones by $\hat{a}$.  
While $A_\mu^a$ give rise to low energy gauge theory, $A_5^{\hat{a}}$ are 4D scalars (transforming as bifundamentals of $SU(3)_D\times SU(3)_c$) and their masses are $\sim {\cal O}(1/R)$ due to quantum corrections from other bulk fields. 

Now we discuss how to break the residual $U(1)$. 
For this purpose, we can have a three-index antisymmetric scalar $\phi_{[ijk]}$ under $SU(6)$. When $i,j,k \in \{1,2,3\}$, then $\phi_{[ijk]}$ transforms as a singlet under both $SU(3)_D$ and $SU(3)_c$, but not under $U(1)$. 
To see this, consider a general set of indices $i,j,k,l$ for which
\begin{align}
	D_\mu \phi_{ijk} \supset A_\mu^a\left[{T^a}_{i}^l \phi_{ljk} + {T^a}_{j}^l \phi_{ilk} + {T^a}_{k}^l \phi_{ijl}\right],
\end{align}
where $T^a$ are various $SU(6)$ generators.
For $i=1, j=2, k=3$, the above becomes,
\begin{align}
	A_\mu^a\left[{T^a}_{1}^1 \phi_{123} + {T^a}_{2}^2 \phi_{123} + {T^a}_{3}^3 \phi_{123}\right].
\end{align}
This implies $\phi_{123}$ is charged under the $U(1)$ since it couples to the diagonal generators. Correspondingly, if $\langle \phi_{123}\rangle \neq 0 $, the $U(1)$ gets broken, leaving only $SU(3)_D\times SU(3)_c$.

In this scenario, the SM Higgs is a singlet as far as orbifolding is concerned and we can put it in the bulk. 
We put SM leptons and quarks on the $y=0$ boundary. 
Since $SU(6)$ is broken into $SU(3)_D\times SU(3)_c$ on this boundary, SM quarks need not fill up a whole multiplet of $SU(6)$, and we take them to be singlets under $SU(3)_D$.
Next, we have to choose parities of SM fermions under ${\cal P}$ and ${\cal P}'$ since the entire Lagrangian must have a definite parity.
Under both ${\cal P}$ and ${\cal P}'$  we take all the SM fermions and SM Higgs to have + parity. 
Then all the SM Yukawa terms are manifestly parity invariant.

Next, we discuss how to generate the intermediate scale $\Lambda \ll M_{\rm GUT}$, which we rely upon for our Higgs portal coupling that allows the dark glueballs to decay. 
We consider vector-like fermions $\chi_L  = (3, 1, 2, -1/2)$ and $\chi_e^c = (\bar{3}, 1, 1, +1)$ under $SU(3)_D \times SU(3)_c \times SU(2)_L \times U(1)_Y$, and their partners, $\chi_L^c  = (\bar{3}, 1, 2, +1/2)$ and $\chi_e = (3, 1, 1, -1)$ located on the $y=0$ boundary. They couple to the Higgs via,
\begin{align}
	y_\chi \chi_L H \chi_e^c + m_{\chi_L} \chi_L \chi_L^c + m_{\chi_e} \chi_e \chi_e^c + {\rm h.c.} \,,
	\label{eq:vector-like}
\end{align}
where $m_{\chi_L}$ and $m_{\chi_e}$ are vector-like mass parameters.
Then, $\chi_L$ and $\chi_e$ mediate a one loop interaction between the $SU(3)_D$ gluons and the Higgs. 
The effective dimension-6 operator may be computed as~\cite{Juknevich:2009gg}
\begin{align}
    \frac{\alpha_D}{6\pi}\frac{y_\chi^2}{m_{\chi_L} m_{\chi_e}} 
    |H|^2 G_{d,\mu \nu} G_d^{\mu\nu} \,.
\end{align}
Therefore the scale $\Lambda$ controlling the glueball decay rate in~\eqref{eq:glueball_decay} corresponds to the masses of the heavy vector-like fermions: $\Lambda^2 / c_6 \sim  m_{\chi_L} m_{\chi_e} / y_\chi^2$. 
Consequently, $\Lambda \ll M_{\rm GUT}$ may be achieved by arranging  vector-like masses $m_{\chi_L} , m_{\chi_e} \ll M_{\rm GUT}$. 

Recall that in the discussion of~\eqref{eq:glueball_dim6} we rely on the $\tilde c_6$ coupling to $|H|^2 G_{d,\mu \nu} \tilde G_{d}^{\mu \nu}$ to induce the decay of the CP-odd glueballs. 
In the theory above, this operator is not generated because the theory is CP conserving. 
However, the theory may be made CP violating by having at least two non-degenerate generations of vector-like fermions, with the associated mass and Yukawa matrices appearing in~\eqref{eq:vector-like} being complex.  
For two generations there is one surviving CP-violating phase that may not be transformed away, while more CP-violating phases survive for a larger number of generations. 
In the presence of at least a single CP-violating phase the $\tilde c_6$ operator appearing in~\eqref{eq:glueball_dim6} is generated, in addition to the $c_6$ operator, as the result of CP violation.

\subsection{Orbifold construction of $SU(8)\rightarrow SU(3)_D \times SU(3)_c \times SU(2)_L \times U(1)_Y$}
\label{sec:orbifold_SU8}

We now describe how the $SU(6)$ group described in the previous subsection can also be unified with $SU(2)_L\times U(1)_Y$ into an $SU(8)$ group.
Since $SU(8)$ has rank 7 and $SU(3)_D \times G_{\rm SM}$ has rank 6, to obtain the above breaking pattern we consider a scalar VEV, such as $\langle \phi_{123}\rangle \neq 0$ in the previous subsection, to reduce the rank.

We first discuss the orbifold parities of the gauge fields. We again consider a $S_1/(Z_2\times Z_2')$ geometry and choose, 
\begin{equation}
    \begin{aligned}
    P = {\rm diag}(-1, -1, -1, +1, +1, +1, +1, +1), \\
    P' = {\rm diag}(-1, -1, -1, -1, -1, -1, +1, +1).
    \end{aligned}
\end{equation}
The choice of $P$ breaks $SU(8)\rightarrow SU(3)_D \times SU(5) \times U(1)_X$. 
On the other hand, $P'$ breaks $SU(8)\rightarrow SU(6) \times SU(2) \times U(1)_Z$. 
Here $U(1)_X$ is generated by ${\rm diag} (r, r, r, s, s, s, s, s)$ with $3r+5s=0,3r^2+5s^2=1/2$, while $U(1)_Z$ is generated by ${\rm diag} (p, p, p, p, p, p, q, q)$ with $6p+2q=0,6p^2+2q^2=1/2$. 
These are the tracelessness and normalization constraints, respectively.
With their combined action, however, the gauge group is broken to 
\begin{align}
    SU(8)\rightarrow SU(3)_D \times SU(3)_c \times SU(2)_L \times U(1)_G \times U(1)_H.
\end{align}
Here we can choose the $U(1)_G$ generator to be ${\rm diag} (r, r, r, s, s, s, t, t)$ with $3r+3s+2t = 0$ (zero trace) and  $(3r^2 + 3s^2+2t^2)=1/2$ (normalized), and 
the $U(1)_H$ generator to be ${\rm diag} (0, 0, 0, p, p, p, q, q)$ with $3p+2q=0$ and $3p^2+2q^2 = 1/2$. 
These conditions determine $p=1/\sqrt{15}$ and $q = -3/(2\sqrt{15})$.

Let us now discuss the embedding of the SM Higgs. 
We put the Higgs in the bulk and in the antifundamental of $SU(8)$, since we can remove the unwanted components by orbifold projection. 
Under $P$, we assume $(+,+,+,+,+,+,+,+)$ parity, while under $P'$, we assume $(-,-,-,-,-,-,+,+)$. This implies only the $SU(2)_L$ doublet has $+$ parity under both $P$ and $P'$, and we can identify the corresponding zero mode as the SM Higgs. 
All the other components are heavy.

Focusing on the SM fermions, we note that we can put them on the $y=0$ boundary since they fit in a multiplet of $SU(5)$, and then we take them as singlets under $SU(3)_D$. 
Next, we need to assign them proper parities such that we can construct Yukawa-invariant terms. 
We choose all the fermions to have + parity under $P$ and $\{+,-,-,+,-\}$ under $P'$, for $q, u^c, d^c, l, e$, respectively. 
Along with the parity requirement on the Higgs, this lets us write appropriate SM Yukawa terms.

To generate the intermediate scale $\Lambda$ that controls the glueball decay rate, we require heavy fermions $\psi_L$ and $\psi_e^c$ on the $y=\pi R/2$ boundary. 
Under the residual $SU(6)\times SU(2)_L \times U(1)_Z$, $\psi_L$ and $\psi_e^c$ have charges $(6, 2, q_1)$ and $(\bar{6}, 1, q_2)$, where $q_1+q_2 = - \sqrt{3}/4$. 
Here the $U(1)_Z$ charge of the SM Higgs, embedded into an $SU(8)$ antifundamental, is taken to be $\sqrt{3}/4$.
We also have vector-like partners $\psi_L^c$ and $\psi_e$ having charges $(\bar{6}, 2, -q_1)$ and $(6, 1, -q_2)$, respectively. 
With these charge assignments, we can write down the Higgs coupling and the vector-like mass terms for the heavy fermions:
\begin{align}
    y_\psi \psi_L H \psi_e^c + m_{\psi_L}\psi_L\psi_L^c + m_{\psi_e}\psi_e \psi_e^c + {\rm h.c.} \,.
\end{align}
Choosing $+$ parity under both $P$ and $P'$ for these fermions makes the above terms parity invariant.
Just as the previous subsection, these heavy fermions mediate an interaction between $SU(3)_D$ and the Higgs and additionally also between $SU(3)_c$ and the Higgs:
\begin{align}
    \frac{\alpha_D}{6\pi}\frac{y_\psi^2}{m_{\psi_L} m_{\psi_e}} |H|^2 G_{d,\mu \nu} G_d^{\mu\nu} + \frac{\alpha_3}{6\pi}\frac{y_\psi^2}{m_{\psi_L} m_{\psi_e}} |H|^2 G_{\mu \nu} G^{\mu\nu}.
\end{align}

\begin{figure}
    \centering
    \includegraphics[width = 0.6\textwidth]{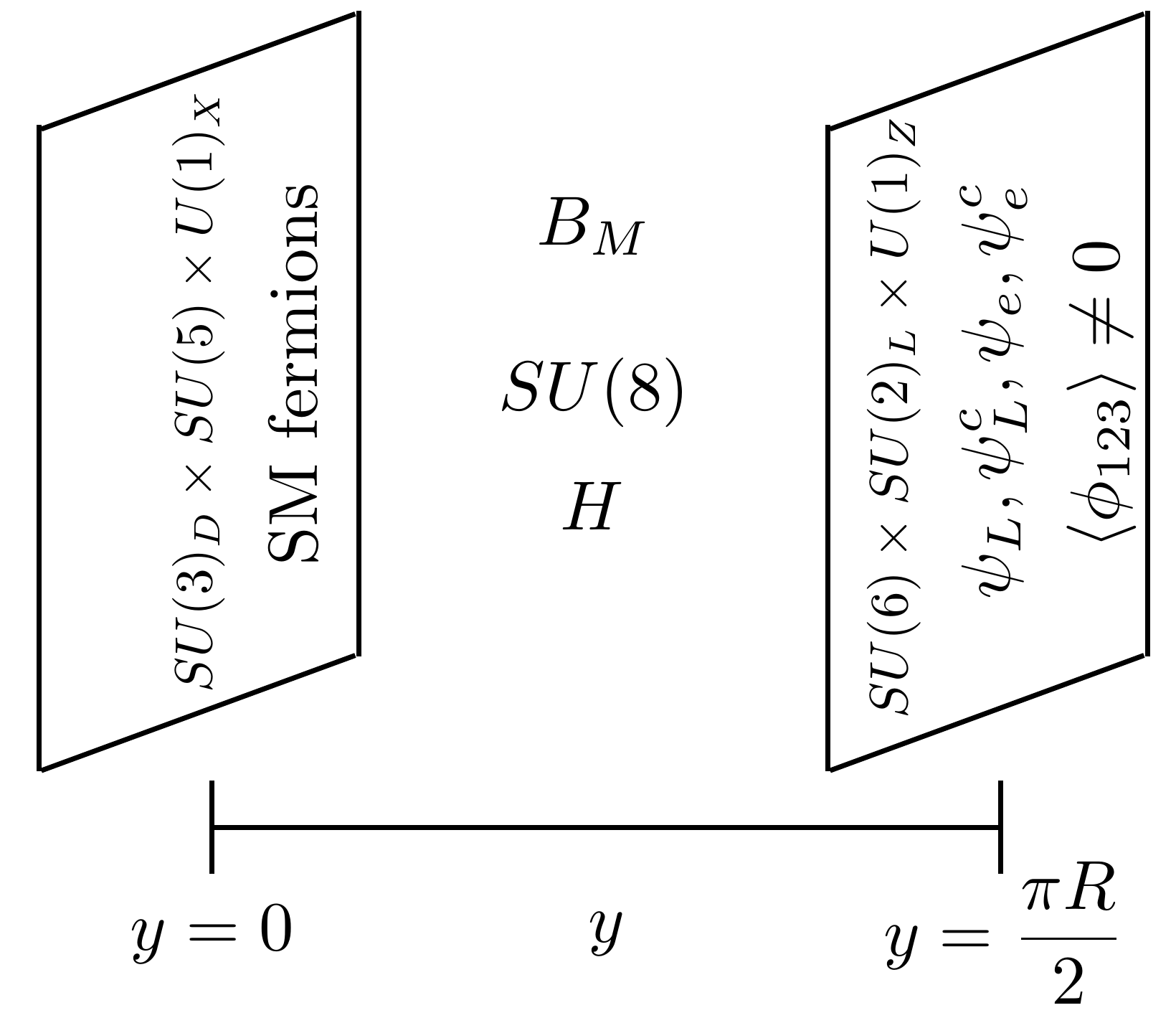}
    \caption{Orbifold Structure for $SU(8)\rightarrow SU(3)_D \times SU(3)_c \times SU(2)_L \times U(1)_Y$ breaking, as described in Sec.~\ref{sec:orbifold_SU8}.  The bulk has an $SU(8)$ gauge symmetry, which is broken by the boundary conditions at $y = 0$ and $y = \pi R / 2$, as indicated, with $y$ the coordinate in the compact fifth dimension. The SM fermions live at the $y = 0$ boundary and do not form full representations of $SU(8)$, since that gauge symmetry is broken at $y = 0$.  The SM Higgs in embedded in an antifundamental of $SU(8)$ in the bulk; in the low energy theory only the $SU(2)_L$ doublet remains light. The heavy vectorlike fermions $\psi_L$, $\psi_e$ and their partners located on the $y = \pi R / 2$ boundary generate the intermediate scale $\Lambda$ that allows the dark glueballs of $SU(3)_D$ to decay to SM Higgs pairs.  The heavy axion arises as the zero mode of the fifth component of a $U(1)$ bulk vector field $B_M$, with the mass of the axion originating primarily from $SU(3)_D$ instantons. 
    }
    \label{fig:5D}
\end{figure}

To break $U(1)_G \times U(1)_H \rightarrow U(1)_Y$, we consider a three-index, totally anti-symmetric scalar of $SU(6)$, $\phi_{[ijk]}$. 
Among its elements, $\phi_{123}$ is a singlet under $SU(3)_D \times SU(3)_c \times SU(2)_L \times U(1)_H$. 
However, it is charged under $U(1)_G$. 
To see this, consider the covariant derivative for general indices $i,j,k,l$ as before,
\begin{align}
    D_\mu \phi_{ijk} \supset	
    A_\mu^a\left[{T^a}_{i}^l \phi_{ljk} + {T^a}_{j}^l \phi_{ilk} + {T^a}_{k}^l \phi_{ijl}\right].
\end{align}
Focusing on $\phi_{123}$ in particular, we see,
\begin{align}
D_\mu \phi_{123} \supset	A_\mu^a\left[{T^a}_{1}^1 \phi_{123} + {T^a}_{2}^2 \phi_{123} + {T^a}_{3}^3 \phi_{123}\right] = A_\mu^a ({T^a}_{1}^1 + {T^a}_{2}^2 + {T^a}_{3}^3) \phi_{123}.
\end{align}
Thus it is charged under only those generators for which $({T^a}_{1}^1 + {T^a}_{2}^2 + {T^a}_{3}^3) \neq 0$. 
Given our choices of $U(1)_G$ and $U(1)_H$, we see that it is charged only under $U(1)_G$. 
Therefore for $\langle \phi_{123}\rangle\neq 0$, the $U(1)_G$ gauge boson gets a mass and $U(1)_H$ survives in the low energy theory. 
Since $U(1)_H$ coincides with the $T_{24}$ generator of $SU(5)$, we can identify this as $U(1)_Y$ along with a multiplicative factor, $Y = c T_{24}$ with $c = -\sqrt{5/3}$. 
This implies $Y = {\rm diag}(0, 0, 0, -1/3, -1/3, -1/3, 1/2, 1/2)$. 
We note that SM fermions need not have any charge under $U(1)_X$ and they inherit their hypercharge from embedding in $SU(5)$, as in the minimal $SU(5)$ model~\cite{Georgi:1974sy}. 
Similarly, the SM Higgs, a part of the antifundamental of $SU(8)$, also obtains the correct hypercharge. 
We summarize the various particle contents and gauge group structure in Fig.~\ref{fig:5D}.

\begin{figure}
    \centering
    \includegraphics[width=0.45\textwidth]{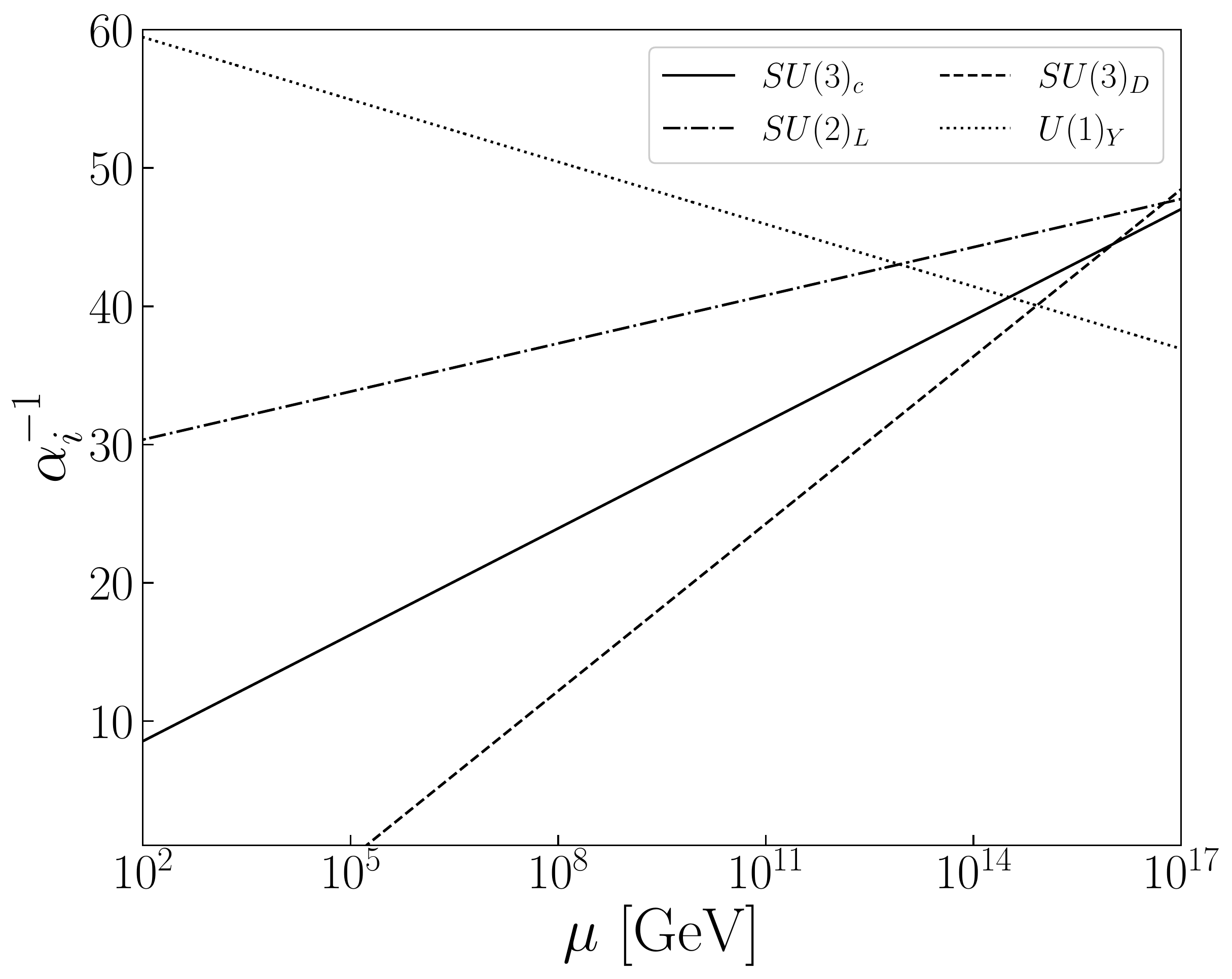}
    \includegraphics[width=0.45\textwidth]{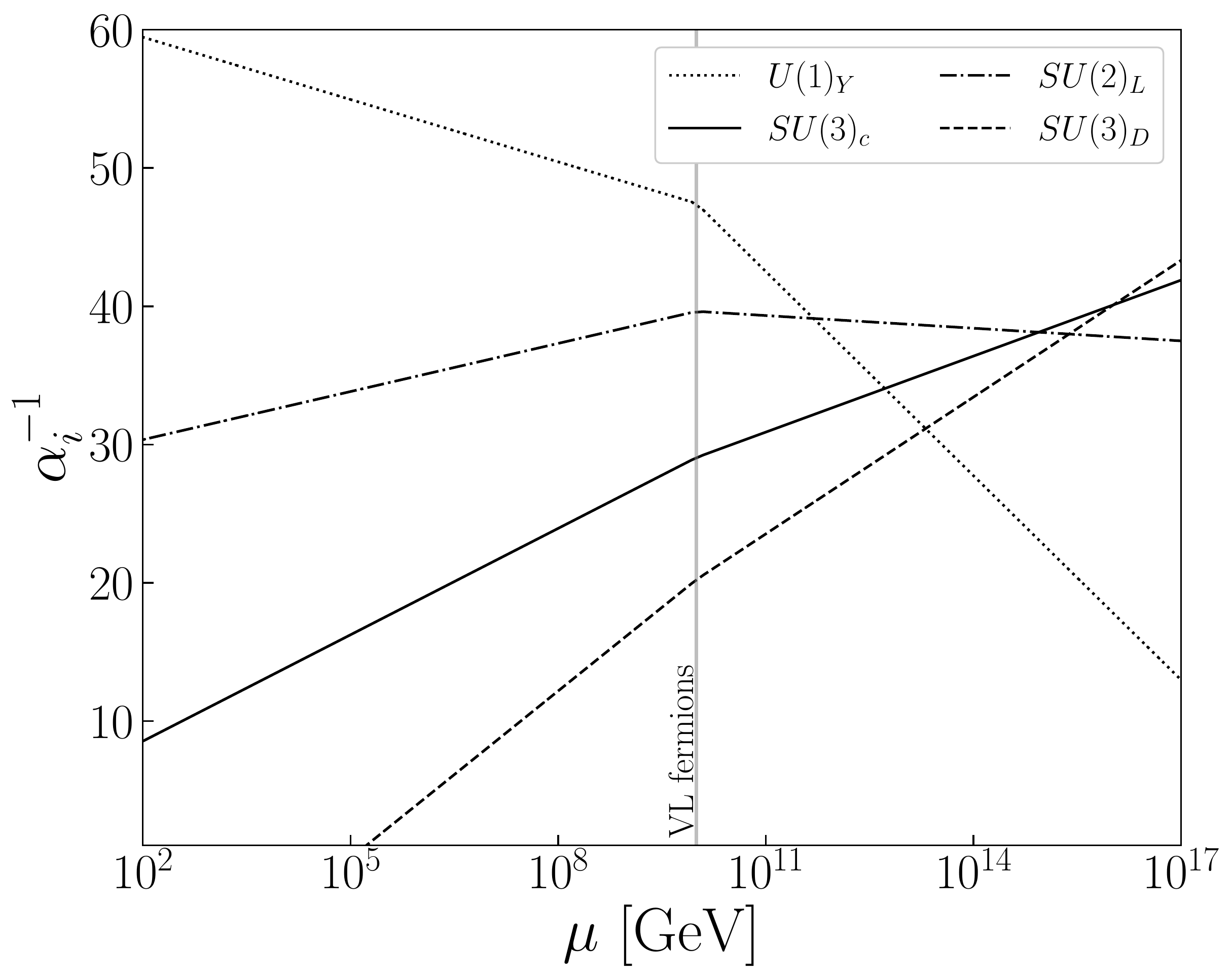}
    \caption{Running of the gauge couplings in the theory where $G_{\rm dark} = SU(3)_D$ unifies with the SM at $\sim$$10^{15}$ GeV and confines at $\sim$$10^5$ GeV.  In the left panel we do not include the vectorlike (VL) fermions in the gauge group running, while in the right panel we do, for the indicated VL fermion mass scale.  As is typically the case for non-supersymmetric GUTs, the gauge-coupling unification is suggestive but not precise.  
    }
    \label{fig:GUT}
\end{figure}

In Fig.~\ref{fig:GUT}, we show the renormalization group evolution of the SM gauge couplings along with that of pure $SU(3)_D$, for a dark confinement scale of $10^5$~GeV. Such a dark confinement scale corresponds to an axion with $m_l \sim 10$~keV and $f_l \sim 10^{15}$~GeV, relevant for the decaying DM parameter space. 
As is well known, the SM gauge couplings do evolve to get close to each other but they do not unify perfectly. 
However, the running of $SU(3)_D$ coupling does indicate unification with $SU(3)_c$ and $SU(2)_L$.
This raises the interesting possibility of achieving a better unification, especially with supersymmetry.

To take into account the effect of vectorlike fermions on the gauge coupling running, we need to know the quantum numbers of the vectorlike fermions under the gauge group $SU(3)_D\times SU(3)_c\times SU(2)_L\times U(1)_Y$. 
We can write the hypercharge operator $Y$ in terms of the $U(1)_Z$ generator $T_Z =(1/\sqrt{48}) {\rm diag}(1, 1, 1, 1, 1, 1, -3, -3)$ and one diagonal generator $T_6 = (1/\sqrt{12}) {\rm diag}(1, 1, 1, -1, -1, -1)$ of $SU(6)$,
\begin{align}
    Y = -\frac{1}{6}\left(\sqrt{48}T_Z - \sqrt{12} T_6 \right).
\end{align}
This gives $Y = {\rm diag}(0, 0, 0, -1/3, -1/3, -1/3, 1/2, 1/2)$. Thus the fermion representation under the bigger group $SU(6)\times SU(2)_L \times U(1)_Z$ splits under $SU(3)_D\times SU(3)_c\times SU(2)_L\times U(1)_Y$ as,
\begin{align}
    (6, 2, q_1) \rightarrow (3, 1, 2, Y_{\psi_L}) + (1, 3, 2, Y_{\psi_L}'),\\
    (\bar{6}, 1, q_2) \rightarrow (\bar{3}, 1, 1, Y_{\psi_e^c}) + (1, \bar{3}, 1, Y_{\psi_e^c}'),
\end{align}
with $Y_{\psi_L} = (-1/6)(\sqrt{48} q_1- \sqrt{12}), Y_{\psi_L}' = (-1/6)(\sqrt{48}q_1+\sqrt{12})$, and $Y_{\psi_e^c} = (-1/6)(\sqrt{48}q_2+\sqrt{12}), Y_{\psi_e^c}' = (-1/6)(\sqrt{48}q_2-\sqrt{12})$. We have $Y_{\psi_L} + Y_{\psi_e^c} = 1/2$, following from $q_1+q_2 = -\sqrt{3}/4$, necessary for the Higgs Yukawa couplings.

\subsection{Axions from extra dimensional gauge fields}

Having discussed the SM sector, we can now include an axion also using the extra dimension. 
We model the axion as the fifth component of a gauge field $U(1)$ in the bulk, following the construction in {\it e.g.}~\cite{Choi:2003wr}. 
We can choose the following parity action on the gauge field,
\begin{equation}
\begin{aligned}
{\cal P}: B_\mu(x,y) &\rightarrow B_\mu(x,-y)  = - B_\mu(x,y),\\
{\cal P}: B_5(x,y) &\rightarrow B_5(x,-y)  = B_5(x,y),\\
\end{aligned}
\end{equation}
with identical action of $\cal P'$ with $y$ replaced by $y'$. 
In other words, while $B_\mu$ has a $--$ parity, $B_5$ has a $++$ parity and only it survives in the low energy theory. 
In the presence of this new gauge field, we can write down a Chern-Simons~(CS) term in the bulk. 
The Lagrangian involving $B_M$ then reads as
\begin{equation}
\begin{split}
\int d^4 x \int_0^{\pi R} dy {}  \Big(\frac{1}{4 g_{5B}^2}B_{MN}B^{MN}  +  \kappa_B \epsilon^{MNPQR}B_M \text{Tr}(F_{NP}F_{QR})\Big).   
\end{split}    
\end{equation}
In the 4D effective theory, this reduces to 
\begin{align}
\frac{\pi R}{2g_{5B}^2} (\partial_\mu B_5)^2 + 2\pi R \kappa_B B_5 G\tilde{G}.
\end{align}
Here, $G$ contains all the $SU(8)$ gauge bosons, which implies that the axion will couple both to the dark $SU(3)$ and to the SM gauge groups. 
Denoting $B_5\equiv a$ and canonically normalizing the kinetic term, we arrive at an axion coupling
\begin{align}
\frac{a}{32\pi^2 f_a}G\tilde{G} \,, \qquad 
f_a \equiv \frac{1}{64\pi^2 \sqrt{\pi R}\kappa_B g_{5B}} \,.
\end{align}
To estimate $f_a$ relative to the unification scale $M_{\rm GUT} \sim 1/R$, we use the relation $\pi R/g_5^2 = 1/g_4^2$ and $4\pi/g_4^2 \approx 25$ at the unification scale, to compute 
\begin{align}
f_a \sim \frac{1}{64\pi^3 \kappa_B} \frac{5 M_{\rm GUT}}{2\sqrt{\pi}}.   
\end{align}
If we suppose that the 5D CS term arises at one loop, such that $\kappa_B \sim \alpha / (4 \pi)$, then numerically $f_a\sim M_{\rm GUT}$.  
Thus, the orbifold model discussed in this section, while by no means unique, contains all of the necessary features needed for the heavy DM axion and baryogenesis stories -- a dark, confining gauge group that unifies with the SM but that contains Higgs portal interactions that allow the dark glueballs to decay, suppressed by an intermediate scale $\Lambda < M_{\rm GUT}$, along with an axion that couples to the SM and to the dark gauge group.

\section{Discussion}
\label{sec:conclu}

In this Article we introduce keV - MeV axions as a decaying DM candidate that may naturally obtain the correct relic abundance through the period of early matter domination brought upon by dark glueballs.  
These glueballs are associated with the dark gauge group whose instantons give rise to the axion mass.  
Such a scenario may naturally arise in an axiverse, where there are multiple axions, in addition to dark gauge groups that decouple from the SM near the GUT scale. 
While such scenarios may emerge in the context of String Theory constructions, which are known to produce decoupled dark gauge groups and axions, we provide an explicit construction in the context of a 5D orbifold theory where the SM and a dark $SU(3)$ unify into a 5D $SU(8)$ theory, which also produces a 4D axion as the zero mode of the fifth component of a 5D gauge field.
We also show that the heavy axions could be responsible for the primordial baryon asymmetry, through the process of spontaneous baryogenesis, and if the dark sector contains multiple confining sub-sectors the correct baryon and DM abundances can simultaneously be produced, as we demonstrate.  The presence of the heavy axions does not spoil the possibility of an additional axion solving the strong {\it CP} problem. 

The clearest signature of heavy axion DM is the decay to two photons, which may be detected by current or near-term $X$-ray and gamma-ray telescopes, as we discuss. 
As illustrated in {\it e.g.} Figs.~\ref{fig:lifetime_plot} and~\ref{fig:tau_bary}, much of the best-motivated parameter space where dark-sector axions may naturally make up the observed DM abundance and also explain the primordial baryon asymmetry could be probed by future instruments, providing strong motivation for missions that increase the reach to the DM lifetime over the keV - MeV energy range.

The dark-sector DM axion cosmology considered in this Article is associated with a period of early matter domination caused by the dark glueballs.  
The fact that low reheat temperatures, near the BBN limit, are favored for mitigating fine tuning of the initial axion misalignment angle may itself lead to observational signatures. 
This is because density perturbations grow linearly during matter-dominated epochs, as opposed to logarithmically during radiation-domination~\cite{Erickcek:2011us,Barenboim:2013gya,Fan:2014zua,Nelson:2018via,Visinelli:2018wza}.  
This implies that small-scale structure could be enhanced because of the period of early matter domination, potentially leading to large numbers of ultra-compact sub-halos that survive until today.  
For reheating temperatures near the BBN bound this implies an enhancement of DM substructure today at masses near Jupiter's mass and below~\cite{Erickcek:2011us}.  
Interestingly, these ultra-compact sub-halos may be directly observable with future Pulsar Timing Array measurements~\cite{Dror:2019twh,Lee:2020wfn} and photometric microlensing surveys~\cite{Dai:2019lud} if $T_{\rm RH} \lesssim 100$ MeV -- GeV, as is the case for most of the parameter space considered in this work.  
It would be interesting to also investigate the possible observational signatures of the ultra-compact mini-halos in the Galactic DM decay morphology.

The period of early matter domination is brought upon by the confining phase transition in the dark non-abelian gauge sector, and depending on the dark gauge group the phase transition could be first order and associated with an efficient production of gravitational waves, see, {\it e.g.}, ~\cite{Schwaller:2015tja, Huang:2020crf, Halverson:2020xpg}. 
The detectability of these gravitational waves at future observatories depends on the efficiency of their production, the temperature of the phase transition, and the amount of subsequent entropy dilution; this would a useful direction to explore in future work.  

In this Article we have not assumed high-scale supersymmetry, except for roughly motivating the gauge couplings that we may expect at the GUT scale. 
Supersymmetry, even if broken at a high scale, would quantitatively and potentially qualitatively modify most of the arguments presented in this work. 
It would be interesting to investigate the supersymmetric completion of the models presented in this work.  

In summary, heavy axions connected to hidden sectors are motivated extensions of the SM that could be responsible for baryogenesis and DM. 
A number of upcoming astrophysical missions should shed light onto their existence, providing strong science motivation for continuing deeper explorations of the cosmos.

\section*{Acknowledgements}

We thank 
Pouya Asadi, Valerie Domcke, Lawrence Hall, Jim Halverson, Keisuke Harigaya, Simon Knapen, Nadav Outmezguine, Nick Rodd, and Raman Sundrum for useful discussions.  
We also thank Pouya Asadi, Valerie Domcke, Jim Halverson, and Nick Rodd for useful comments on the manuscript.
J.W.F. was supported by a Pappalardo Fellowship.  B.R.S. was supported  in  part  by  the  DOE  Early Career  Grant  DESC0019225. S.K., B.R.S., and Y.S. were supported in part by a grant from the United States-Israel Binational Science Foundation
(BSF No.~2020300), Jerusalem, Israel. 
Y.S. is also supported by grants from the ISF (No.~482/20), NSF-BSF (No.~2018683) and by the Azrieli foundation. 
J.W.F. and S.K. thank the Mainz Institute of Theoretical Physics of the Cluster of Excellence PRISMA+ (Project ID 39083149) for its hospitality while this work was in progress.
This research used resources from the Lawrencium computational cluster provided by the IT Division at the Lawrence Berkeley National Laboratory, supported by the Director, Office of Science, and Office of Basic Energy Sciences, of the U.S. Department of Energy under Contract No.  DE-AC02-05CH11231.

\appendix

\section{Sensitivity projections for future gamma ray observatories}
\label{sec:astro}
In this Appendix we give the details for deriving the sensitivity projections for AMEGO. Projections for other missions in the $\sim$ MeV energy range can be obtained in a similar way.

For an observation of an on-sky region $\Sigma$ for duration $T$ using an instrument with energy-dependent effective area $\mathcal{E}$, the expected number of observed photons produced by decay of an axion with mass $m_a$ to two photons with energy $m_a / 2$ is given by 
\begin{equation}
    N(m_a, \tau_a) = \frac{\mathcal{D}_{\Sigma} \mathcal{E}(m_a/2) T}{2 \pi m_a \tau_{a}} \,,
\end{equation}
where $\tau_a$ is the lifetime for axion decay to photons, and $\mathcal{D}_\Sigma$ is the DM line-of-sight density integrated over the region of interest. 
Assuming the axion comprises all the DM, we consider axion decays in the Milky Way halo in the vicinity of the GC with $\Sigma$ defined by $|b|,\, |l| \leq 5^\circ$.
We take the Milky Way DM density profile to be described by an NFW profile \cite{Navarro:1995iw, Navarro:1996gj}, though more motivated and better constrained modeling choices for the Milky Way DM density profile may be possible in the future through improved simulation and observational efforts \cite{deSalas:2020hbh}. We take our NFW profile to have DM density $0.4\,\mathrm{GeV}/\mathrm{cm}^3$ in the solar neighborhood at $r_\odot = 8.23\,\mathrm{kpc}$ from the galactic center (GC) and a scale radius of $r_s = 20\,\mathrm{kpc}$ \cite{deSalas:2020hbh, 2022arXiv220412551L}. The integrated line-of-sight density is then calculated by
\begin{gather}
    \rho(r) =\rho_\odot \frac{  r_\odot (r_s+ r_\odot)^2}{r (r_s+r)^2} \\
    \mathcal{D}_\Sigma = \int_\Sigma d\Omega \int ds \rho_{a}(s, \Omega) \,,
\end{gather}
such that $\mathcal{D}_\Sigma \approx 4 \times 10^{24}\, \mathrm{MeV}/\mathrm{cm}^2$. We assume the region of interest $\Sigma$ is observed for $T = 1\,\mathrm{yr}$ and adopt AMEGO's projected energy-dependent effective area. For the energy range relevant for our axion DM scenario, AMEGO will observe incident photons as Compton scattering events with two different classifications: tracked and untracked. The Tracked Compton (TC) and Untracked Compton (UC) event classifications cover complementary energy ranges, with differences in effective area and energy resolution \cite{Kierans:2020otl}. In projecting AMEGO sensitivities, we independently consider both event types. 

To estimate our statistical power in constraining an axion decay line, we calculate the expected number of background photons contributed by astrophysical processes within the energy range over which the signal appears using flux spectra for bremsstrahlung, inverse-Compton, and $\pi^0$ emission in the $|b|,\, |l| \leq 5^\circ$ region developed in \cite{Bartels:2017dpb} using  the cosmic ray modeling code \texttt{GALPROP}\cite{Strong:1998pw}. Since AMEGO will not resolve the decay line-width, which has relative width of $\Delta E / E \approx 10^{-3}$, the relevant energy range is the instrumental energy resolution, which is roughly $\Delta E /E \approx 5\%$, evaluated at $E = m_a / 2$. The number of background photons $N_B$ is then given by \begin{equation}
N_B(m_a) = \int_\Sigma d\Omega\frac{d\Phi}{dE d\Omega} \mathcal{E}(m_a/2) T \Delta E(m_a/2) \,,
\end{equation}
where $\mathcal{E}$ and $\Delta E(E)$ are the energy-dependent effective area and energy-resolution appropriate chosen for the tracked or untracked event classifications. From $N_B$ and $N(m_a, \tau_a)$, the expected 95$^\mathrm{th}$ percentile upper limit on $\tau_a$ can be determined in the gaussian limit relevant to these projections by solving $N(m_a, \tau_ a) \approx 1.6 N_B(m_a)$ \cite{Cowan:2010js}.  Note that we neglect systematic uncertainties, which may be important, especially at low energies where the photon counts are the highest~\cite{Bartels:2017dpb}, to show the maximal possible reach of the instruments from statistical uncertainties alone. 

The projected sensitivity of AMEGO for tracked and untracked event types are presented in Fig.~\ref{fig:lifetime_plot}, labeled `AMEGO TC' and `AMEGO UC', respectively. Note that we have neglected the finite angular resolution of the instrument, though this is a small correction as the angular resolution is comparable to or less than the extent of our region of interest and since the line-of-sight DM density is not sharply varying outside the very inner GC.

\section{Rate formulae for lepton-asymmetry generating operators}
\label{sec:rates}
In this Appendix we summarize the interaction rates $\Gamma_\alpha$ relevant for lepton asymmetry generation via the Boltzman equation~\eqref{eq:bary_boltz}. 
These interactions, if active, are responsible for maintaining chemical equilibrium between different SM species.
We recall that $\alpha$ runs over weak sphaleron ($W$), strong sphaleron ($S$), tau Yukawa ($\tau$), top Yukawa ($t$), bottom Yukawa ($b$), and the Weinberg
operator for the first two generations ($W_{12}$) and the third generation ($W_3$). 
\begin{figure}[!htb]
    \centering
    \includegraphics[width=0.49 \textwidth]{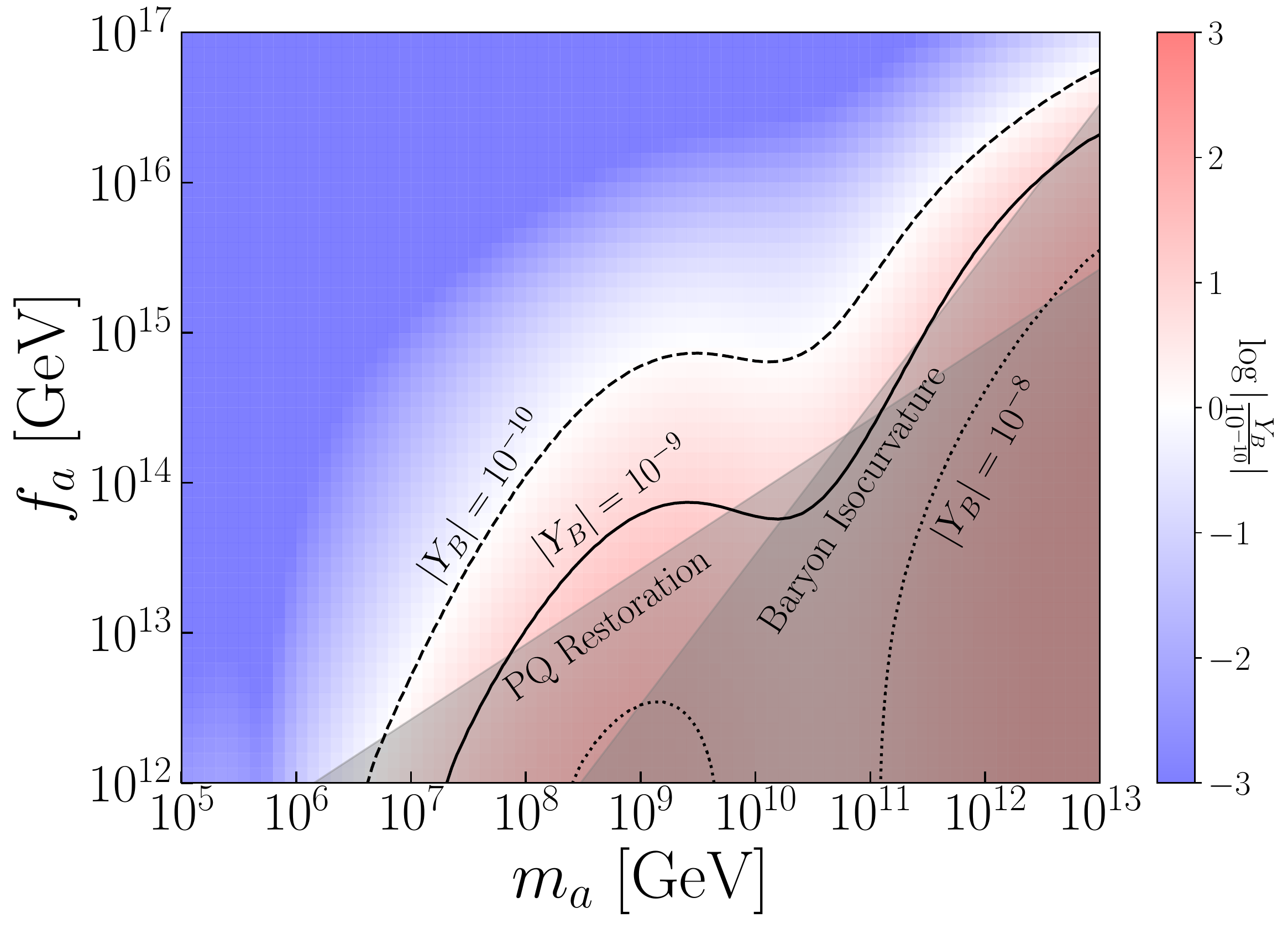}
    \includegraphics[width=0.49 \textwidth]{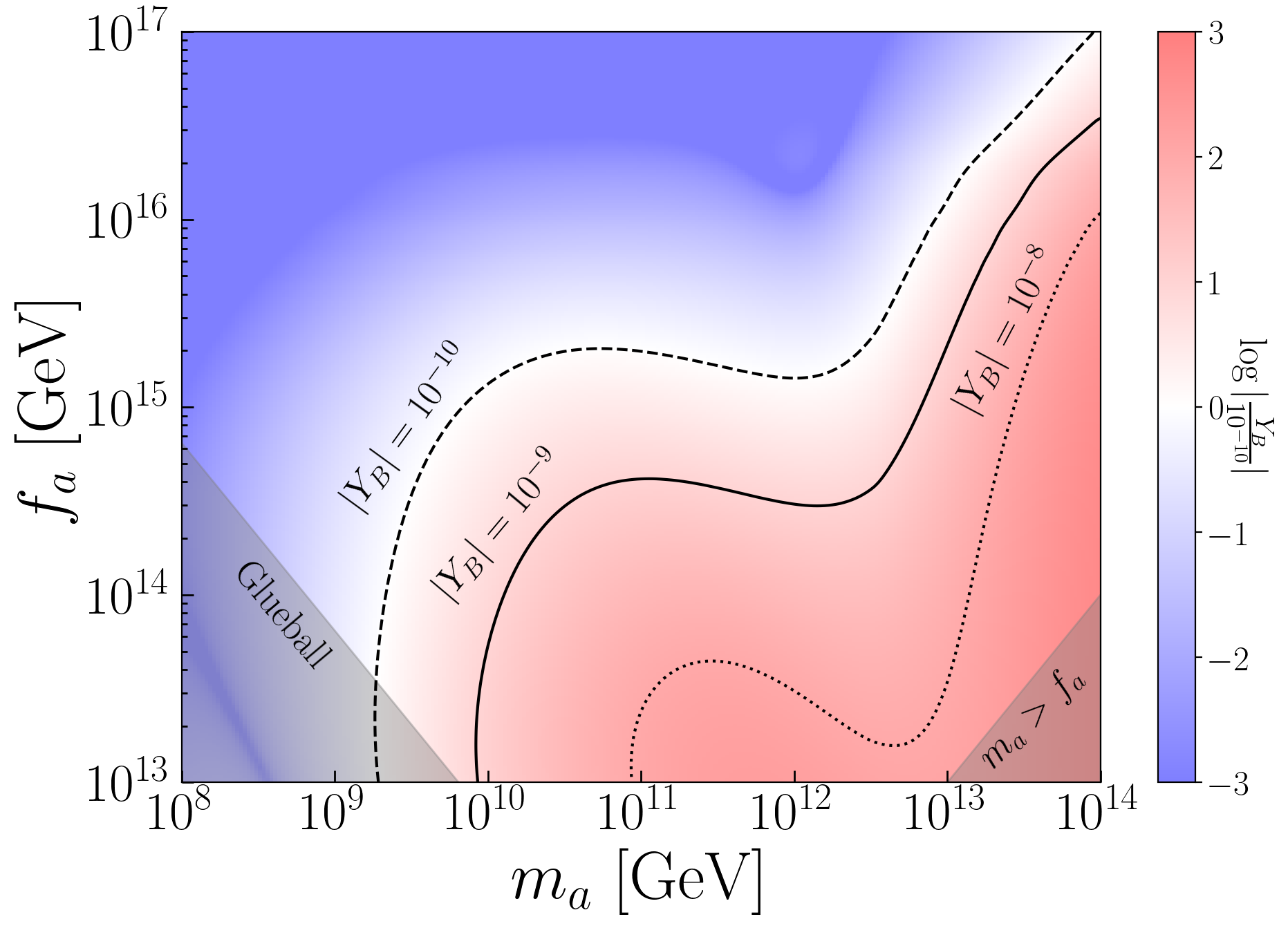}
    \caption{As in Fig.~\ref{fig:heavy_ax_T_ind} but with $c_{aG} = 1$, $c_{af} = 0$ instead of $c_{aG} = c_{af} = 1$ for the heavy axion. This implies that the axion does not couple to fermions and that suppresses the induced baryon asymmetry. 
    }
    \label{fig:ax_gauge_only_T_ind}
\end{figure}

Since the distinction between the first two generations is immaterial at high temperatures, they can be combined into a single species, and a single Weinberg operator interaction $W_{12}$ can describe them.
\paragraph{Strong and weak sphaleron.}
The sphaleron rates in gauge theories can be obtained from, {\it e.g.},~\cite{Moore:2010jd}
\begin{align}
    \Gamma_{W} = 3\kappa_{W}\alpha_2^5 T,\\
    \Gamma_{S} = 3\kappa_{S}\alpha_s^5 T.
\end{align}
We take $\kappa_{W}\sim 24$ and $\kappa_{S}\sim 270$~\cite{Domcke:2020kcp} as relevant for asymmetry generation at high temperatures, $T\gtrsim 10^{12}$~GeV.
\paragraph{Yukawa couplings.}
Given the sizes of the Yukawa couplings, here only the third generation fermions is relevant,
\begin{align}
    \Gamma_\tau = 6 \kappa_\tau y_\tau^2 T,\\
    \Gamma_t = 6 \kappa_t y_t^2 T, \\
    \Gamma_b = 6 \kappa_b y_b^2 T.
\end{align}
Here we take $\kappa_\tau \simeq 1.7\times 10^{-3}, \kappa_t \simeq \kappa_b \simeq 10^{-2}$~\cite{Domcke:2020kcp, Garbrecht:2014kda}.
\paragraph{Weinberg operator.}
This can be estimated as~\cite{Domcke:2020kcp}
\begin{align}
    \Gamma_{W_{12}} = 2\Gamma_{W_3} = 12\kappa_W \frac{m_\nu^2 T^3}{v^4},
\end{align}
with $\kappa_W \sim 3\times 10^{-3}$, $m_\nu = 0.05$~eV and $v = 174$~GeV.
\section{Alternate axion-matter coupling choice for spontaneous baryogenesis}
\label{app:alt}

Recall that in constructing Fig.~\ref{fig:heavy_ax_T_ind} for the parameter space that produces the correct baryon asymmetry we make the choice $c_{aG} = c_{af} = 1$.  In this Appendix we consider the alternate choice $c_{aG} = 1$, $c_{af} = 0$, which implies that the axion couples to gauge fields but has no direct coupling to fermions (see~\eqref{eq:axion_couplings}).  In Fig.~\ref{fig:ax_gauge_only_T_ind} we show the analogue of Fig.~\ref{fig:heavy_ax_T_ind} for this alternate choice of heavy axion couplings.  Note that the baryon asymmetries are generically suppressed relative to in the case where the axion couples at tree level to fermions; this is mostly because the axion-top coupling is the most efficient operator for baryon production, so removing this operator suppresses the baryon asymmetry production.

\bibliographystyle{JHEP}
\bibliography{refs}

\end{document}